\documentclass[11pt]{article}
\usepackage{axodraw2,pix}
\usepackage{epsfig}
\usepackage{amsfonts}
\usepackage{amsmath}
\usepackage{bbm,bm}
\usepackage{cite}
\allowdisplaybreaks[1]

\def\dralgo{{\tt DRalgo}}
\newcommand{\DRalgoVersion}{{\tt 1.0}}

\usepackage{color}
\usepackage{algorithm,algpseudocode}

\algnewcommand{\algorithmiccall}{{\bf Call}}
\algdef{SE}[FOR]{Call}{EndCall}[1]
  {\algorithmiccall\ #1\ \{}
  {\}} 

\usepackage{dirtree}

\def\backtick{\char18}
\usepackage{listings}
\usepackage[utf8]{inputenc}
\usepackage{tikz}
\usetikzlibrary{shapes,arrows,shadows,automata,positioning}
\newdimen\nodeDist
\usepackage[
  colorlinks=true,
  linkcolor=black,
  citecolor=blue,
  filecolor=black,
  urlcolor=black,
  breaklinks=true
  ]{hyperref}

\renewcommand{\vec}[1]{{\bf #1}}

\newcommand{\gammaE}{{\gamma_\rmii{E}}}

\newcommand{\Lamd}{\bmu_{3}}
\newcommand{\LamD}{\bmu}
\newcommand\MSbar{$\overline{\rm MS}$}
\renewcommand{\tr}{{\rm Tr\,}}

\newcommand{\nf}{n_{\rm f}}

\newcommand{\Tc}{T_{\rm c}}

\newcommand{\T}{\rmii{$T$}}
\newcommand{\Yl}{Y_{\ell}}
\newcommand{\Yq}{Y_{q}}
\newcommand{\Ye}{Y_{e}}
\newcommand{\Yu}{Y_{u}}
\newcommand{\Yd}{Y_{d}}

\newcommand{\mD}{m_\rmii{D}}

\newcommand{\bmu}{\bar\mu}

\def\lsi{\raise0.3ex\hbox{$<$\kern-0.75em\raise-1.1ex\hbox{$\sim$}}}
\def\gsi{\raise0.3ex\hbox{$>$\kern-0.75em\raise-1.1ex\hbox{$\sim$}}}

\renewcommand{\nn}{\nonumber \\}
\renewcommand{\rmi}[1]{{\mbox{\scriptsize #1}}}
\newcommand{\rmii}[1]{{\mbox{\tiny\rm{#1}}}}

\newcommand{\re}{\mathop{\mbox{Re}}}
\newcommand{\im}{\mathop{\mbox{Im}}}

\newcommand{\Tint}[1]{{\hbox{$\sum$}\!\!\!\!\!\!\!\int\,}_{\!\!\!\!\raise-0.9ex\hbox{$\scriptstyle{#1}$}}}
\newcommand{\Tinti}[1]{{{\Sigma}\!\!\!\!\raise0.3ex\hbox{$\int$}_\rmii{${#1}$}}}
\newcommand{\Tintip}[1]{{{\Sigma'}\!\!\!\!\!\raise0.3ex\hbox{$\int$}_\rmii{${#1}$}}}

\newcommand{\bsl}[1]{\,\slash\!\!\!\!{#1}\,}

\newcommand{\deltabar}{\raise-0.02em\hbox{$\bar{}$}\hspace*{-0.8mm}{\delta}}

\newcommand*\ol[1]{\overline{#1}}
\newcommand{\eps}{\epsilon}

\newcommand{\vev}{vacuum expectation value}

\def\scfc{0.7}  
\def\phgt{21}   
\def\pwc{21}    
\def\pwcb{31.5} 
\def\pwcc{42} 
\newcommand{\PIC}[4]{\;\parbox[c]{#2 pt}{\begin{picture}(#2,#3)(0,0)
\SetWidth{1.0}\SetScale{#4} #1 \end{picture}}\;}
\renewcommand{\pic}[1]{\PIC{#1}{\pwc}{\phgt}{\scfc}}
\renewcommand{\picb}[1]{\PIC{#1}{\pwcb}{\phgt}{\scfc}}
\renewcommand{\picc}[1]{\PIC{#1}{\pwcc}{\phgt}{\scfc}}

\makeatletter \@addtoreset{equation}{section} \makeatother
\renewcommand{\theequation}{\arabic{section}.\arabic{equation}}
\makeatletter
\renewcommand\section{\@startsection{section}{1}{\z@}%
  {-5.5ex \@plus -1ex \@minus -.2ex}
  {2.3ex \@plus.2ex}%
  {\normalfont\large\bfseries}}
\renewcommand\subsection{\@startsection{subsection}{2}{\z@}%
  {-3.25ex\@plus -1ex \@minus -.2ex}%
  {1.5ex \@plus .2ex}%
  {\normalfont\normalsize\bfseries}}
\renewcommand\thesection{\@arabic\c@section}
\renewcommand\thesubsection{\thesection.\@arabic\c@subsection}
\renewcommand{\@seccntformat}[1]{%
  \csname the#1\endcsname.\hspace{1.0em}}
\makeatother

\begin{document}

\flushbottom

\begin{titlepage}

\begin{minipage}{.5\textwidth}
\flushright
\includegraphics[height=1.5cm]{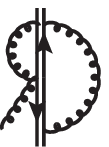}
\end{minipage}
\begin{minipage}{.5\textwidth}
\flushright
HIP-2022-11/TH\\
NORDITA 2022-030 \\
\end{minipage}

\begin{centering}

\vfill


{\Large{\bf
\dralgo{}:
  a package for effective field theory approach for\\
  thermal phase transitions 
}}

\vspace{0.8cm}

\renewcommand{\thefootnote}{\fnsymbol{footnote}}
Andreas Ekstedt$^{\rm a,b,c,}$%
\footnotemark[1],
Philipp Schicho$^{\rm d,}$%
\footnotemark[2],
and Tuomas V.~I.~Tenkanen$^{\rm e,f,g,}$%
\footnotemark[3]

\vspace{0.8cm}

$^\rmi{a}$%
{\em
Department of Physics and Astronomy, Uppsala University,\\
P.O.~Box 256, SE-751 05 Uppsala, Sweden\\}
\vspace{0.3cm}

$^\rmi{b}$%
{\em
II. Institute of Theoretical Physics, Universität Hamburg, D-22761, Hamburg, Germany\\}
\vspace{0.3cm}

$^\rmi{c}$%
{\em
Deutsches Elektronen-Synchrotron DESY, Notkestr. 85, 22607 Hamburg, Germany\\}
\vspace{0.3cm}

$^\rmi{d}$%
{\em
Department of Physics and Helsinki Institute of Physics,\\
P.O.~Box 64, FI-00014 University of Helsinki,
Finland\\}
\vspace{0.3cm}

$^\rmi{e}$%
{\em
Nordita,
KTH Royal Institute of Technology and Stockholm University,\\
Roslagstullsbacken 23,
SE-106 91 Stockholm,
Sweden\\}
\vspace{0.3cm}

$^\rmi{f}$%
{\em
Tsung-Dao Lee Institute \& School of Physics and Astronomy,\\
Shanghai Jiao Tong University, Shanghai 200240, China\\}
\vspace{0.3cm}

$^\rmi{g}$%
{\em
Shanghai Key Laboratory for Particle Physics and Cosmology, Key Laboratory for Particle Astrophysics and Cosmology (MOE), Shanghai Jiao Tong University, Shanghai 200240, China\\}

\vspace*{0.8cm}

\mbox{\bf Abstract}

\end{centering}

\vspace*{0.3cm}

\noindent
\dralgo{} is an algorithmic implementation 
that constructs an effective,
dimensionally reduced,
high-temperature field theory for generic models.
The corresponding Mathematica package automatically performs the matching to
next-to-leading order.
This includes
two-loop thermal corrections to scalar and Debye masses as well as 
one-loop thermal corrections to couplings.
\dralgo{} also allows for integrating out additional heavy scalars.
Along the way, the package provides
leading-order beta functions for general gauge-charges and fermion-families; both in
the fundamental and in the effective theory.
Finally, the package computes the finite-temperature effective potential within the effective theory.
The article explains the theory of the underlying algorithm
while introducing the software on a pedagogical level.

\vfill
\end{titlepage}

\tableofcontents
\renewcommand{\thefootnote}{\fnsymbol{footnote}}
\footnotetext[1]{andreas.ekstedt@desy.de}
\footnotetext[2]{philipp.schicho@helsinki.fi}
\footnotetext[3]{tuomas.tenkanen@su.se}
\clearpage

\renewcommand{\thefootnote}{\arabic{footnote}}
\setcounter{footnote}{0}

%

\clearpage
%
\section*{Program summary}
{\em Program title}: 
Dimensional Reduction algorithm (\dralgo{})
\\%
{\em Version}:
{\tt \DRalgoVersion{}}
\\%
{\em Developer's repository link}:
\url{https://github.com/DR-algo/DRalgo}
\\%
{\em Licensing provisions}:
GNU General Public License 3 (GPLv3)
\\%
{\em Programming languages}:
{\tt Mathematica} 
\\%
{\em External routines/libraries}:
{\tt GroupMath}~\cite{Fonseca:2020vke}
\\%
{\em Nature of problem}:
Construction of
high-temperature effective field theories for
beyond the Standard Model physics.
\\%
{\em Solution method}:
Matching of $n$-point correlation functions using
tensor-notation of couplings~\cite{%
  Martin:2017lqn,Martin:2018emo,Machacek:1984zw,Machacek:1983fi,Machacek:1983tz}.
\\%
{\em Restrictions}:
{\tt Mathematica} version 12 or above.
\\%
{\em Operating system}:
macOS~11 and higher,
Linux Ubuntu 18.04.
\clearpage

%
\section{Introduction}
\label{sec:intro}

The origin of Baryon asymmetry~\cite{Canetti:2012zc} in the universe remains obscure.
As such,
much powder has been spent throughout past decades to find
a sound explanation of this asymmetry~\cite{%
  Elor:2022hpa,Shaposhnikov:1987tw,Shaposhnikov:1986jp,Kuzmin:1985mm}.
Amongst the suggested mechanisms,
the one based on the electroweak phase transition --
electroweak baryogenesis -- stands out.
While the Standard Model (SM) has no strong first-order transition on its own~\cite{Kajantie:1995kf,Gurtler:1997hr,Kajantie:1996mn},
its extensions can contain myriads of them.
For new field content to trigger a strong first-order phase transition,
their masses have to be around the electroweak scale, and
their interaction with the SM Higgs cannot be too feeble.
Beyond-the-Standard-Model (BSM) theories
that exhibit such transitions provide a direct target for many
future-generation colliders~\cite{Ramsey-Musolf:2019lsf,Friedrich:2022cak}.
Furthermore, thermal phase transitions can, perchance, generate
gravitational waves that are observable by next-generation space-based detectors such as
LISA~\cite{LISA:2017pwj},
DECIGO~\cite{Kawamura:2006up},
Taiji~\cite{Guo:2018npi} and
BBO~\cite{Harry:2006fi}.
These waves might open a fresh window into the early universe --
and
the underlying quantum field theory.

The interplay between BSM phenomenology and gravitational waves is among 
the most actively studied topics in
the high-energy-physics literature~\cite{%
  deVries:2017ncy,Dorsch:2018pat,Hall:2019ank,Baldes:2018nel,Chala:2018opy,Croon:2020cgk,Gould:2021dzl,Alves:2018jsw,Niemi:2021qvp,Bell:2020hnr,
  Niemi:2020hto}.
This interplay
requires solid understanding of thermodynamic properties of different models.
It has been long known that determining thermodynamics in theories with non-Abelian gauge fields is challenging because of
the Linde problem~\cite{Linde:1980ts}.
In short, there are non-perturbative effects arising from  massless vector-bosons: these infrared (IR) effects can only be captured by
lattice simulations~\cite{Braaten:1994na}.
Despite this, leading-order perturbation theory is frequently used as an approximation. 
However, it has been pointed out that such leading-order studies contain large theoretical uncertainties, due to slow convergence of
perturbation theory~\cite{Arnold:1992rz,Farakos:1994kx,Croon:2020cgk,Gould:2021oba}.

Dimensional reduction~\cite{Ginsparg:1980ef,Appelquist:1981vg} offers a way to overcome these challenges.
In this framework,
ultraviolet (UV) modes --
related to non-zero Matsubara modes in the imaginary time formalism --
are integrated out. 
The resulting effective field theory (EFT)~\cite{Kajantie:1995dw,Braaten:1995cm} describes
IR, or
long wavelength zero modes, living in three spatial dimensions (3d)
(cf.\ also~\cite{Kajantie:1997hn,Andersen:1997ba,Andersen:1996dh}).
The 3d EFT can be simulated on a lattice and hence by-pass
the Linde problem~\cite{Braaten:1994na,Farakos:1994xh}. 
Dimensional reduction can also be viewed as
a systematic scheme for thermal resummations, used purely within the realm of perturbation theory \cite{Farakos:1994kx}.
To leading order, dimensional reduction is equivalent to resummation via thermal masses.
Concretely, the effective potential, 
that describes the free energy of thermal plasma, 
is commonly given in a schematic form~\cite{Arnold:1992rz}
\begin{align}
\label{eq:veff-1loop}
  V_{\rmi{eff}} \simeq
      V_{T=0}
    + V_{\T}
    + V_{\rmi{daisy}}
  \;. 
\end{align}
Here,
$V_{T=0}$ is the Coleman-Weinberg zero-temperature contribution,
which is often augmented with the effective potential at one-loop level.
Namely,
$V_{\T}$ is the thermal correction at one-loop
(commonly in the high-temperature expansion) and 
\begin{align}
\label{eq:daisy}
V_{\rmi{daisy}} \simeq
    - \frac{T}{12\pi} \Big[
        (m^2 + \Pi_\T)^\frac{3}{2}
      - (m^2)^{\frac{3}{2}}
    \Big]
    \;. 
\end{align}
Here $m^2$ represents a field-dependent mass and $\Pi_\T \sim g^2 T^2$ is a one-loop thermal correction. It is the first term in eq.~\eqref{eq:daisy} that implements the resummation.%
\footnote{
  The second $\sim T m^3$ term in eq.~\eqref{eq:daisy} 
  removes double counting;
  the same contribution is included $V_\T$ but with unresummed mass.
  This term is related to the Matsubara zero-mode contribution.
}
This resummed mass, for zero Matsubara modes arises from thermal corrections due to
non-zero Matsubara modes.
Physically, this corresponds to thermal screening:
UV modes in the plasma screen IR modes.
In eq.~\eqref{eq:veff-1loop} all contributions are evaluated at one-loop order, which means that only masses are resummed.
However, at high temperatures, the perturbative expansion is controlled by $g$, rather than $g^2$ as in a zero-temperature loop expansion.
Dimensional reduction makes it possible to systematically include
higher order resummations, in particular two-loop thermal masses and one-loop resummation of couplings and fields
which are also affected by thermal screening. 
Hence, dimensional reduction can be used to include important 
next-to-leading order thermal effects.
In particular, renormalization-group (RG) improvements require two-loop computations
at high temperatures \cite{Arnold:1992rz,Gould:2021oba}; these are straightforward to implement in the 3d EFT approach \cite{Farakos:1994kx}.

To date there are sundry studies using dimensional reduction in
electroweak theories~\cite{%
  Andersen:1996dh,Farakos:1994kx,Kajantie:1995dw,Andersen:1997ba,Kajantie:1997hn,
  Gould:2021dzl,Croon:2020cgk,Niemi:2021qvp,Niemi:2020hto,Andersen:2017ika,
  Niemi:2018asa,Hirvonen:2022jba,Rajantie:1997pr}.
For similar studies in hot QCD, see~\cite{Andersen:2015eoa,Andersen:2002ey}.
Still, many models remain to be studied within the dimensional-reduction framework~\cite{%
  Baum:2020vfl,Ivanov:2014doa,BhupalDev:2018xya,Appelquist:2002mw}. 

Along with this article, we launch the Mathematica package \dralgo{} to implement
dimensional reduction automatically for user-defined models.
This package calculates parameters in the 3d EFT.
For example, \dralgo{} calculates
two-loop thermal masses,
effective couplings, and
beta functions.
The package also allows users to integrate out
heavy scalars and
temporal (Debye) scalars.
Furthermore, \dralgo{} can be used to compute
the two-loop thermal effective potential within the effective theory. 
Subsequently,
the 3d EFT matching relations can be implemented to lattice Monte Carlo simulations.
The algorithm~\ref{alg:dralgo} illustrates this pipeline.
\begin{algorithm}
\caption{
\label{alg:dralgo}
  \dralgo{}
  algorithm outline
  in alignment with fig.~2 of~\cite{Schicho:2021gca}.
  The use of lattice resources is indicated as an optional path.
  The functions
  {\tt PerformDRhard[]},
  {\tt PerformDRsoft[]} and
  {\tt CalculatePotentialUS[]}
  are part of \dralgo{}.
}
\begin{algorithmic}
\State {\bf Input:}
  Four-dimensional Lagrangian $\mathcal{L}_{\rmii{4d}}$ with 
  parameters $\{c_{1},\dots,c_{n}\}$,
  temperature $T$,
  physical parameters,
  heavy masses $\bm{M}$
\State {\bf Start:}
  Initialize model 
\Call{ {\tt PerformDRhard[]}}
  \For{{\bf all}  $c_{i} \in \{c_{1},\dots,c_{n}\}$}
    \State
      Compute 4-dimensional $\beta$-functions $\beta(c_{i})$
    \State
      Compute $c_{i,\rmi{3d}}(T,\bm{M})$ by integrating out non-zero Matsubara modes
  \EndFor
  \State 
    Compute thermal (Debye) masses $m_{\rmii{D},i}(T,\bm{M})$
  \State 
    Compute couplings that involve temporal vector fields
\EndCall
\State {\bf Output/Input:}
Three-dimensional soft Lagrangian
$\mathcal{L}_{\rmii{3d}}$ with 
parameters $\{c_{1,\rmi{3d}},\dots,c_{n,\rmi{3d}}\}$
\Call{{\tt PerformDRsoft[]}}
  \For{{\bf all}  $c_{i,\rmi{3d}} \in \{c_{1,\rmi{3d}},\dots,c_{n,\rmi{3d}}\}$}
    \State
      Compute 3-dimensional $\beta$-functions $\beta_{\rmii{3d}}(c_{i,\rmi{3d}})$
    \State
      Compute $\bar{c}_{i,\rmi{3d}}(T,c_{1,\rmi{3d}},\dots,c_{n,\rmi{3d}})$
      by integrating out
      massive temporal scalars
  \EndFor
\EndCall
\State {\bf Output:}
Three-dimensional ultrasoft Lagrangian
$\bar{\mathcal{L}}_{\rmii{3d}}$ with 
  parameters $\{\bar{c}_{1,\rmi{3d}},\dots,\bar{c}_{n,\rmi{3d}}\}$
\If{Lattice resources}
  \State
    Compute lattice continuum relations to construct
    $\mathcal{L}^{\rmii{lattice}}_{\rmii{3d}}$
  \State
    Monte Carlo simulation
\Else
  \Call{{\tt CalculatePotentialUS[]}}
    \State
      Compute effective potential
      $V^{\rmii{3d}}_{\rmii{eff}}(m_{i,\rmi{3d}}^2$)
      up to two-loops 
  \EndCall
\EndIf
\State
  Compute thermodynamic parameters
  $\Tc$, $L/\Tc^4$ 
\end{algorithmic}
\end{algorithm}
This package can be applied to models previously studied in the
electroweak and dark-sector phase-transition literature.   

The remainder of this article is organised as follows.
Section~\ref{sec:prologue} briefly introduces
dimensional reduction and matching relations.
Section~\ref{sec:speedrun} explicates the front-end of the package,
its installation, and gives a tutorial based on the Abelian-Higgs model.
Section~\ref{sec:theory} reviews theory and computations in
the back-end of the package.
Section~\ref{sec:BSM} illustrates additional features by implementing
a two-Higgs doublet model. 
In section~\ref{sec:outlook},
we discuss future prospects and possible updates of the package.
Finally, appendix~\ref{sec:thermo}
displays the computation of thermodynamical observables using
the output of \dralgo{}
for the Abelian-Higgs model.

\subsection{Prologue: dimensional reduction at next-to-leading-order}
\label{sec:prologue}
At leading order (LO)
scalars obtain familiar one-loop thermal masses, while
temporal (longitudinal) scalars get Debye masses.
At next-to-leading order (NLO) couplings are resummed, and
scalars receive two-loop thermal masses.
\dralgo{} also calculates two-loop Debye masses.
Corrections at NLO are particularly important due to large logarithms.
The idea with dimensional reduction is to render some logarithms ($\ln\frac{\bmu}{T}$) small by 
matching at a high energy ($\bmu\sim T$).
The remaining logarithms can then be resummed by
RG-evolution within the EFT.

Phase transitions often occur below the Debye-mass scale.
In such cases temporal scalars can also be integrated out.
Thus effectively removing large logarithms of the form $\ln\frac{\mD}{\bmu}$,
where $\mD$ is a Debye mass.
This second EFT is said to live at the ultrasoft scale,
which is characterised by energies of $\mathcal{O}(g^2 T)$.
In summary,%
\footnote{
  Some literature~\cite{Kajantie:1995dw} interchangeably refers to
  the hard scale as superheavy,
  the soft scale as heavy, and
  the ultrasoft scale as light.
}
the hard scale corresponds to energies $E\sim T$;
the soft scale corresponds to energies $E\sim g T$; and
the ultrasoft scale corresponds to energies $E\sim g^2 T$.

To illustrate next-to-leading order dimensional reduction,
we consider a schematic model with
scalar mass parameter $\mu^2$,
scalar quartic coupling $\lambda$, and
gauge coupling $g$. 
Given the power counting
$\mu^2 \sim g^2 T^2$,
$\lambda \sim g^2$,
the matching of the mass parameter is
\begin{align}
\bar{\mu}^2_3 &=
  \underset{\mathcal{O}(g^2)}{
    \fbox{$
      \colorbox{gray!10}{$
        \overset{\text{tree-level}}{\mu^2}
        $}
    + \colorbox{gray!10}{$
      \overset{\text{1-loop}}{\# g^2 T^2}
      $}
    $}
  }
  + \underset{\mathcal{O}(g^4)}{
    \fbox{$
      \colorbox{gray!10}{$
      \overset{\text{1-loop}}{\# g^2\mu^2}
        $}
    + \colorbox{gray!10}{$
      \overset{\text{2-loop}}{\# g^4 T^2}
      $}
    $}
    }
  + \mathcal{O}(g^6)
  \nn[2mm] &
  + \underset{\mathcal{O}(g^3)}{
    \fbox{$
    \colorbox{gray!10}{$
    \overset{\text{1-loop}}{\# g^2\mD}
    $}
    $}
  }
  + \underset{\mathcal{O}(g^4)}{
    \fbox{$\colorbox{gray!10}{$
    \overset{\text{2-loop}}{\# g^4}
    $}$}
  }
  + \mathcal{O}(g^5)
  \;,
\end{align}
where
the first line (with even powers of $g$) results from the first step, and
the second line (with odd power of $g$) from second step of the dimensional reduction.
In practice, {\em full} $\mathcal{O}(g^4)$ contributions are included.
Going to higher orders, requires
a three-loop computation for both steps of the dimensional reduction. 
The situation is similar for the coupling:
\begin{align}
\bar{\lambda}_3 &=
    \underset{\mathcal{O}(g^2)}{
    \fbox{$\colorbox{gray!10}{$
      \overset{\text{tree-level}}{T \lambda \vphantom{g^4}}
    $}$}
    }
  + \underset{\mathcal{O}(g^4)}{
    \fbox{$\colorbox{gray!10}{$
    \overset{\text{1-loop}}{\# g^{4}}
    $}$}
  }
  + \mathcal{O}(g^6)
  \nn[2mm] &
  + \underset{\mathcal{O}(g^3)}{
    \fbox{$\colorbox{gray!10}{$
    \overset{\text{1-loop}}{\# \frac{g^4}{\mD^{ }}}
    $}$}
  }
  + \underset{\mathcal{O}(g^4)}{
    \fbox{$\colorbox{gray!10}{$
    \overset{\text{2-loop}}{\# \frac{g^6}{\mD^{2}}}
    $}$}
  }
  + \mathcal{O}(g^5)
  \;. 
\end{align}
In practice for the coupling, $\mathcal{O}(g^4)$ pieces are neglected since
their numerical effect is small (despite being formally of the same order).
These contributions arise during the second step of
the dimensional reduction at two-loop.
Pursuing higher order, requires
a two-loop computation for the first step, and
a three-loop computation for the second step of the dimensional reduction. 

In perturbation theory
the definition of 3d EFT parameters are accompanied with
a perturbative computation of 
the effective potential~\cite{Farakos:1994kx}:
\begin{align}
V^{\text{3d}}_{\text{eff}} =     
    \underbrace{V^{\rmi{3d}}_{\rmi{tree}}}_{\mathcal{O}(g^2)}
  + \underbrace{V^{\rmi{3d}}_{\rmi{1-loop}}}_{\mathcal{O}(g^3)}
  + \underbrace{V^{\rmi{3d}}_{\rmi{2-loop}}}_{\mathcal{O}(g^4)}
  + \mathcal{O}(g^5)
  \;.
\end{align}
Instead of expanding the result in terms of 4d parameters of the parent theory,  
resummed couplings and masses are kept along.
This improves the overall convergence as the result is less sensitive to
the renormalization scale~\cite{Laine:2006cp,Ghiglieri:2020dpq}.
The order $\mathcal{O}(g^5)$ requires a computation at
three-loop level~\cite{Rajantie:1996np,Moller:2012chx,Laine:2018lgj}.

\subsubsection*{Dimensional reduction from a zero-temperature EFT perspective}

It is instructive to consider the same physics from a standard EFT perspective.
In particular, for equilibrium observables we can, analogous to
Kaluza-Klein theories~\cite{Overduin:1997sri},
view thermal corrections as an infinite tower of heavy particles.
To see the connection with dimensional reduction,
we consider a theory with
one light and
one heavy scalar in Euclidean spacetime:
\begin{align}
  \mathcal{L}=
    \frac{1}{2}(\partial_\mu \phi)^2
  + \frac{1}{2}(\partial_\mu \Phi)^2
  + \frac{1}{2}m^2 \phi^2
  + \frac{1}{2}M^2\Phi^2
  + \frac{1}{4}\kappa  \phi^2 \Phi^2
  + \frac{1}{4!}\lambda  \phi^4
  \;.
\end{align}
Imagine now that there exists a hierarchy
$M^2\gg m^2$. 
This is a well-known situation, and when calculating scattering processes one encounters large corrections to $m^2$ scaling as
$\kappa M^2$ --
analogous to the hierarchy problem.
As well as large logarithms in the form
$\ln\frac{\bmu^2}{M^2}$ or
$\ln\frac{\bmu^2}{m^2}$.
One of these logarithms will then be large regardless the choice of
the RG-scale $\bmu$.

Using thermal masses is equivalent to using a resummation%
\footnote{
  This can also be viewed as a temperature-dependent renormalization $\delta_\T m^2$ such that $m^2 = m^2 + \delta_\T m^2$~\cite{Niemi:2021qvp}.
}
$m^2 \to m^2+ a \kappa M^2$.
This resummation does, however,
not solve our problems of large logarithms.
Therefore, calculations, even with thermal masses, are sensitive to the RG-scale.
The situation is much improved by using an EFT.
Therein,
the logarithms
$\ln\frac{\bmu^2}{M^2}$ are rendered small by matching at the scale
$\bmu_\rmii{Match}\sim M$, and
the remaining logarithms
$\ln\frac{\bmu^2}{m^2}$
are taken care of by RG-evolution in the low-energy theory~\cite{Cohen:2019wxr}.

Once the scalar field $\Phi$ has been integrated out,
the resulting EFT is of the form
\begin{align}
\mathcal{L}_\text{eff}=
    \frac{1}{2}(\partial_\mu \phi)^2
  + \frac{1}{2}m^2_\rmi{eff}\,\phi^2
  + \frac{1}{4!}\lambda_\rmi{eff}^{ }\,\phi^4
  \;.
\end{align}
To leading order
$m_\rmi{eff}^{2} = m^2 +a \kappa M^2$, and
$\lambda_\rmi{eff}^{ }=\lambda$.
While at NLO
\begin{align}
  \delta m_\text{eff}^2 &= \kappa^2M^2\Bigl(b+c \ln\frac{\bmu^2_\rmii{Match}}{M^2}\Bigr)
  \;,\\
  \delta \lambda_\rmi{eff} &= \kappa^2\Bigl(d+e \ln\frac{\bmu^2_\rmii{Match}}{M^2}\Bigr)
  \;,
\end{align}
where
$b,c,d,e$ are numerical coefficients.
It is precisely these kind of corrections that appear in the dimensionally reduced theory.

Finally, to do calculations in the IR, we should run the couplings within the EFT from $\bmu_\rmii{Match}\sim M$ down to
$\bmu_\rmii{IR}\sim m_\rmi{eff}$.
In this way, all large logarithms are eliminated and
perturbative convergence is improved.

In the case of dimensional reduction,
we have an infinite tower of heavy particles with masses
$M_n^2=(2\pi n T)^2$ for all integers $n\neq 0$.%
\footnote{
Fermions have masses of $M_n=(2n + 1) \pi T$.
}
Hence,
the high-temperature matching scale is
$\mu_\rmii{Match}\sim T$ and
the low-energy scale is
$\mu_\rmii{IR}^2\sim g^2 T^2$.
Since we are describing equilibrium (static) processes,
there is no time dependence --
the degrees of freedom of the EFT are static and live in three dimensions.

%
\section{Installation and running}
\label{sec:speedrun}
This section explains how to install \dralgo{} and presents
a tutorial based on the Abelian-Higgs model. 

\subsection{Installation}
The current version of \dralgo{}, \DRalgoVersion{}, is installed by
placing all the source files either in
the applications folder
{\tt Mathematica/Applications}
or
running Mathematica from the package root directory {\tt ./DRalgo}.

\begin{figure}[t]
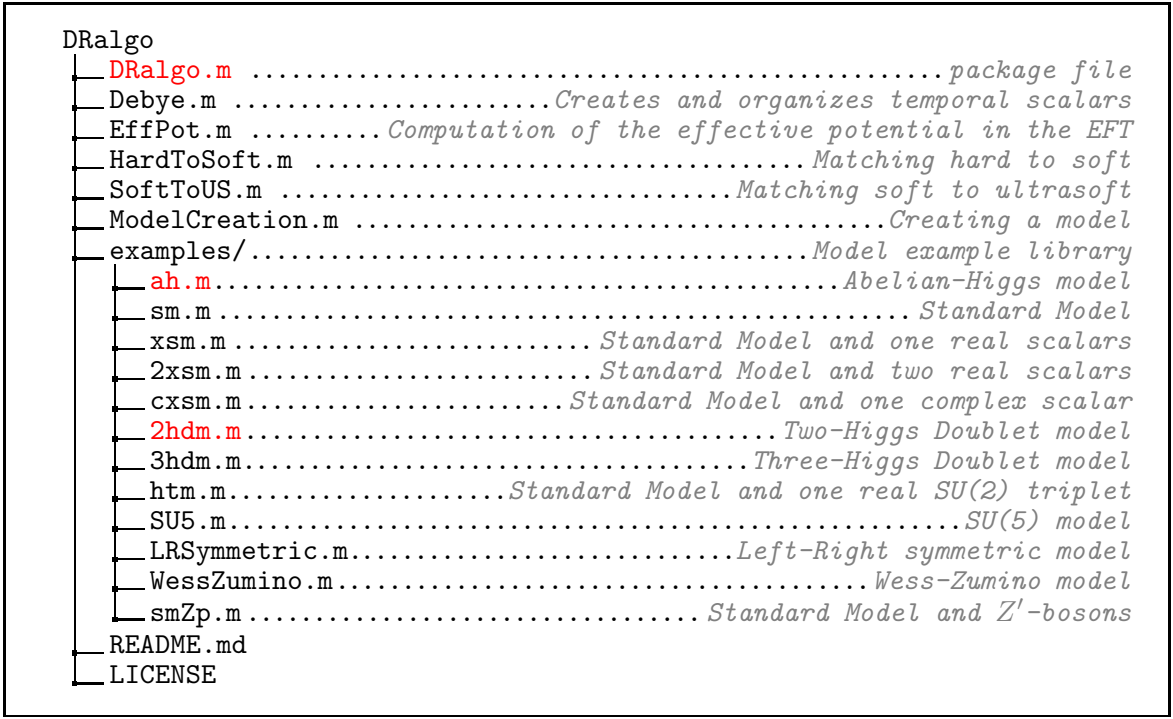

\centering
\framebox[\textwidth]{%
\begin{minipage}{0.95\textwidth}
  \vspace{2mm}
  \dirtree{%
    .1 DRalgo.
    .2 {\color{red} DRalgo.m}
    \DTcomment{package file}.
    .2 Debye.m
    \DTcomment{Creates and organizes temporal scalars}.
    .2 EffPot.m
    \DTcomment{Computation of the effective potential in the EFT}.
    .2 HardToSoft.m
    \DTcomment{Matching hard to soft}.
    .2 SoftToUS.m
    \DTcomment{Matching soft to ultrasoft}.
    .2 ModelCreation.m
    \DTcomment{Creating a model}.
    .2 examples/\DTcomment{Model example library}.
    .3 {\color{red} ah.m}\DTcomment{Abelian-Higgs model}.
    .3 sm.m\DTcomment{Standard Model}.
    .3 xsm.m\DTcomment{Standard Model and one real scalars}.
    .3 2xsm.m\DTcomment{Standard Model and two real scalars}.
    .3 cxsm.m\DTcomment{Standard Model and one complex scalar}.
    .3 {\color{red} 2hdm.m}\DTcomment{Two-Higgs Doublet model}.
    .3 3hdm.m\DTcomment{Three-Higgs Doublet model}.
    .3 htm.m\DTcomment{Standard Model and one real SU(2) triplet}.
    .3 SU5.m\DTcomment{SU(5) model}.
    .3 LRSymmetric.m\DTcomment{Left-Right symmetric model}.
    .3 WessZumino.m\DTcomment{Wess-Zumino model}.
    .3 smZp.m\DTcomment{Standard Model and $Z'$-bosons}.
    .2 README.md.
    .2 LICENSE.
  }
  \vspace{2mm}
\end{minipage}
}
\caption{%
  Outline of the structure of \dralgo{}.
  The most relevant files for this article are highlighted.
  Specific models are discussed in
  sec.~\ref{sec:speedrun} for
  {\tt ah.m} and
  sec.~\ref{sec:BSM} for
  {\tt 2hdm.m}.
}
\label{fig:DirectoryTree}
\end{figure}

The required source files are outlined in fig.~\ref{fig:DirectoryTree}.
To create model files, \dralgo{} uses functions from
{\tt GroupMath}~\cite{Fonseca:2020vke};
see ibid.\ for its installation.
This tutorial uses {\tt GroupMath} Version {\tt 1.1.2} but
{\tt GroupMath} is not required for the dimensional reduction itself.
If a model file is already available,
it can be loaded independently of {\tt GroupMath} (cf.\ sec.~\ref{sec:save}).
Since {\tt GroupMath} is an external package, any use of
the model-creation features in \dralgo{} should be accompanied by
a citation of~\cite{Fonseca:2020vke}.

To load \dralgo{} from the Mathematica applications folder
{\tt Mathematica/Applications},
the following commands need to be executed:
\begin{lstlisting}[language=Mathematica]
SetDirectory[NotebookDirectory[]];
$LoadGroupMath=True;
<<DRalgo@!\backtick@!
\end{lstlisting}
Subsequently, we demonstrate 
the definition of models,
the dimensional reduction, and
the calculation of the two-loop effective potential.
All steps are mirrored in the accompanying
{\tt ./examples/ah.m}.

\subsection{Model implementation}

As an example, we consider
the Abelian-Higgs model with the Euclidean Lagrangian 
\begin{align}
\label{eq:L:ah}
&  \mathcal{L} =
     \frac{1}{4}F_{\mu \nu}F^{\mu \nu}
    + \left(D^\mu \phi\right)^\dagger\left(D_\mu\phi\right)
    + V(\phi,\phi^\dagger)
  \;, \\
&V(\phi,\phi^\dagger) =
      m^2 \phi^{\dagger}\phi
    + \lambda\bigl(\phi^\dagger\phi\bigr)^2
  \;,
\end{align}
where
the covariant derivative is
$D_\mu = \partial_\mu - i g_1 Y_\phi A_\mu$ and 
the corresponding field-strength tensor is 
$F_{\mu \nu} =\partial_\mu A_\nu-\partial_\nu A_\mu$.
Here, the field
$A_\mu$ is an $\mathrm{U}(1)$ gauge field with gauge coupling $g_1$ and
$\phi$ is a complex scalar field charged under $\mathrm{U}(1)$ with general charge $Y_\phi$.

The zero, or temporal, component of the vector field
receives a thermally induced mass in the 3d theory.
This temporal scalar transforms as a singlet under $\mathrm{U}(1)$ and has
a Lagrangian
\begin{align}
\mathcal{L}_{\rmi{temporal}} &=
    \frac{1}{2} (\partial_i A_0)^2
  + \frac{1}{2} {\tt \mu sqU1} A^2_0
  + \frac{1}{4!} \lambda\texttt{VLL[1]} A^4_0
  + \frac{1}{2} \lambda\texttt{VL[1]} A^2_0 \phi^\dagger \phi
  \;,
\end{align}
where indices $i\in \{1,\dots,d\}$ are spatial.
Here,
{\tt $\mu$sqU1} is the temporal-scalar Debye mass (squared),
{\tt $\lambda$VLL[1]} is the self-interaction coupling, and 
{\tt $\lambda$VL[1]} is the portal coupling to $\phi$.
The notation above is aligned with the systematic notation of \dralgo{}.
For earlier studies of this model in the literature, we refer to~\cite{%
  Andersen:1996dh,Andersen:1997ba,Kajantie:1997hn,Hirvonen:2021zej,Karjalainen:1996rk}.

To create a model in \dralgo{}, we have to specify the gauge group and the scalar representation.
In \dralgo{} this is done via
\begin{lstlisting}[language=Mathematica,mathescape=true]
Group={"U1"};
CouplingName={g1};
RepAdjoint={0};
Higgs={{Y$\phi$},"C"};
RepScalar={Higgs}; 
RepFermion={};
\end{lstlisting}
Here,
the adjoint (vector) representation is trivial for the Abelian Higgs model.
In the definition of the Higgs scalar
{\verb!{{Y!$\phi$\verb!},"C<R>"}!},
$Y_\phi$ denotes the gauge charge.
One additional argument is passed depending on if the scalar has
a real ({\tt R}) or
a complex ({\tt C}) representation.
In this simple example, fermions are absent and
the corresponding bracket is empty. 
While not relevant for this model, note that fermions are never part of the 3d EFT Lagrangian as they are integrated out
during the first step of the dimensional reduction.
The model-input from the gauge sector is generated with the command
\begin{lstlisting}[language=Mathematica,mathescape=true]
{gvvv,gvff,gvss,$\lambda$1,$\lambda$3,$\lambda$4,$\mu$ij,$\mu$IJ,$\mu$IJC,Ysff,YsffC}=
    AllocateTensors[Group,RepAdjoint,CouplingName,RepFermion,RepScalar];
\end{lstlisting}
The tensors listed above have the following correspondence
\begin{center}
\begin{tabular}{ r l } 
  {\tt gvvv} & structure constants \\ 
  {\tt gvff} & vector-fermion trilinear couplings \\
  {\tt gvss} & vector-scalar trilinear couplings \\ 
  {\tt $\lambda$1} & scalar tadpole couplings \\
  {\tt $\lambda$3} & cubic couplings \\
  {\tt $\lambda$4} & quartic couplings \\
  {\tt $\mu$ij} & scalar-mass matrix \\
  {\tt $\mu$IJ}, 
  {\tt $\mu$IJC} & fermion-mass matrices \\ 
  {\tt Ysff}, 
  {\tt YsffC} & Yukawa couplings 
\end{tabular}
\end{center}
and their purpose is described in depth in sec.~\ref{sec:theory}. 
By default all these tensors, except for the gauge ones, are empty.
While their order on the left hand side of
{\tt AllocateTensors} is fixed,
the above naming of {\verb!{gvvv,...,YsffC}!}
is arbitrary.

The next task is to specify the scalar potential.
Let us start with the mass matrix.
There is only one allowed term: $\phi \phi^\dagger$.
This term is selected via
\begin{lstlisting}[language=Mathematica,mathescape=true]
InputInv={{1,1},{True,False}};
MassTerm1=CreateInvariant[Group,RepScalar,InputInv][[1]]//Simplify;
VMass=msq*MassTerm1; (* corresponds to a term $\frac{1}{2} m^2 \phi \phi^{\dagger}$ *)
$\mu$ij=GradMass[VMass];
\end{lstlisting}
To understand the syntax, recall that we defined our scalar as
{\verb!RepScalar={Higgs}!}.
Then
{\verb!{1,1}!} specifies that we are after a term with two ({\tt RepScalar[[1]]}) $\phi$ fields.
For general models, 
the index {\tt 1} can be replaced by any index in {\tt RepScalar} as defined by the user. 
In addition,
{\verb!{True,False}!} specifies that we are after
the $\phi \phi^\dagger$ term and
conversely
{\verb!{True,True}!} corresponds to
a $\phi \phi$ term.
The latter term, however,
is not allowed in this example due to gauge invariance.
Since this example exhibits merely
one $\mathrm{U}(1)$ invariant combination of $\phi$ and $\phi^\dagger$,
we specify {\tt [[1]]} in the {\tt CreateInvariant} output.

For the quartic tensor only one possible term arises, namely
$(\phi \phi^\dagger)^2$.
We already created a
$(\phi \phi^\dagger)$ term above,
which can be reused to find the quartic:
\begin{lstlisting}[language=Mathematica,mathescape=true]
VQuartic=$\lambda$*MassTerm1^2;
$\lambda$4=GradQuartic[VQuartic]
\end{lstlisting}
To see how
quartic and mass terms are composed of scalar-field components,
one can inspect the variables
{\tt VMass} and
{\tt VQuartic}.

This completes the model implementation, and we now have all ingredients to do
the dimensional-reduction step.

\subsection{Running the dimensional-reduction algorithm}

First we need to load the model with the command
\begin{lstlisting}[language=Mathematica,mathescape=true]
ImportModelDRalgo[Group,gvvv,gvff,gvss,$\lambda$1,$\lambda$3,$\lambda$4,$\mu$ij,$\mu$IJ,$\mu$IJC,Ysff,YsffC,Verbose->False];
\end{lstlisting}
Since the gauge group was previously defined,
the corresponding Debye masses are automatically generated and named.
The option {\tt Verbose->False} disables all progress messages. 

The actual dimensional reduction is performed with the command
\begin{lstlisting}[language=Mathematica,mathescape=true]
PerformDRhard[];
\end{lstlisting}
This calculates all
thermal masses,
effective couplings,
$\beta$-functions and
anomalous dimensions.
For example, gauge and scalar couplings are given by
\begin{lstlisting}[language=Mathematica,mathescape=true]
PrintCouplings[]
$
\Bigl\{
  \text{g13d}^2 \to
    g^2_1  T
  - \frac{g^4_1\text{Lb}T Y^2_\phi}{48\pi^2}
  ,\;
  \lambda\text{3d} \to
  \frac{
      T(
      g^4_1  Y^4_\phi (2-3\text{Lb})
    + 6 g^2_1  Y^2_\phi \lambda \text{Lb}
    + 2 \lambda (8\pi^2 - 5 \lambda\text{Lb})
    )}{16 \pi ^2}
  \Bigr\}
$
\end{lstlisting}
where the output is a replacement rule.
The variables
{\tt Lb} and
{\tt Lf} are dependent on 
the 4d-renormalization scale ($\LamD$) 
and
the temperature ($T$)~\cite{Kajantie:1995dw,Farakos:1994kx}.
They are 
\begin{align}
\label{eq:lblf}
{\tt Lb} &= \ln\Bigl(\frac{\bmu^2}{T^2}\Bigr)
+ 2\gammaE
- 2\ln(4\pi)
\;,\quad 
{\tt Lf} = {\tt Lb} + 4\ln(2)
\;,
\end{align}
where
$\gammaE$ is the Euler-Mascheroni constant. 
Further information about the functional basis
of the effective parameters is collected in sec.~\ref{sec:basis}.
The corresponding definitions of constants,
along with other shorthand notations,
are shown with the command
\begin{lstlisting}[language=Mathematica,mathescape=true]
PrintConstants[]
\end{lstlisting}

The above RG-scale $\LamD$ is the hard matching scale, and should be chosen
as $\LamD \sim T$ to avoid large logarithms.%
\footnote{
  In \dralgo{} the renormalization scales are denoted by
  ${\tt \mu} = \LamD$ for the hard scale and
  ${\tt \mu 3} = \Lamd$ for the soft scale. 
}
Effective mass parameters are divided into scalar and temporal-scalars.
The scalar masses are given by
\begin{lstlisting}[language=Mathematica,mathescape=true]
PrintScalarMass["LO"]
PrintScalarMass["NLO"]
\end{lstlisting}
Here,
the LO command prints tree-level masses together with one-loop thermal contributions; the NLO command prints two-loop thermal masses.
By default these NLO masses contain
the running from the matching scale ($\LamD$) to an arbitrary 3d scale ($\Lamd$).%
\footnote{
  There are different ways to express the NLO masses.
  We refer to sec.~\ref{sec:basis} for further details.
}
The resulting Debye masses are displayed by
\begin{lstlisting}[language=Mathematica,mathescape=true]
PrintDebyeMass["LO"]
PrintDebyeMass["NLO"]
\end{lstlisting}
Mixed temporal-scalar couplings (to NLO)
are printed with
\begin{lstlisting}[language=Mathematica,mathescape=true]
PrintTemporalScalarCouplings[]
\end{lstlisting}
Such temporal-scalar couplings are denoted as  
$\lambda${\verb!VLL[a]!} for $V^4$ couplings,
$\lambda${\verb!VVSL[a]!} for $V^2 S$ couplings and
$\lambda${\verb!VL[a]!} for $V^2S^2$ couplings,
where
$S$ represents scalar fields and 
$V$ temporal scalar fields. 
The index
`{\tt a}'
systematically labels all linearly independent terms.

The coefficient of the unit operator~\cite{Braaten:1995cm},
or the hard-mode contribution to the symmetric-phase pressure, is given by
\begin{lstlisting}[language=Mathematica,mathescape=true]
PrintPressure["LO"]
PrintPressure["NLO"]
PrintPressure["NNLO"]
\end{lstlisting}
Here,
LO corresponds to one-loop,
NLO to two-loop, and
NNLO to three-loop level.
The LO result describes the pressure of an ideal gas, and is given by
$P_{\rmii{LO}}(T) = N \frac{\pi^2}{90} T^4$, where
$N$ represents the number of degrees of freedom.
For the Abelian Higgs model
$N=4$ with
2 from the photon and
2 from the complex scalar. 

\dralgo{} also provides anomalous dimensions and beta functions in
the parent 4d theory; see sec.~\ref{sec:BetaFunc}. 
Beta functions are printed with the command
\begin{lstlisting}[language=Mathematica,mathescape=true]
BetaFunctions4D[]
\end{lstlisting}

Next, to find anomalous dimensions we first need to specify for which particles we want them.
As \dralgo{} stores everything in tensor form, we need to find the positions of all particles
\begin{lstlisting}[language=Mathematica,mathescape=true]
PosScalar=PrintScalarRepPositions[];
\end{lstlisting}
To determine the anomalous dimension of e.g.\ {\tt ScalarRep[[1]]},
we would write
\begin{lstlisting}[language=Mathematica,mathescape=true]
AnomDim4D["S",{PosScalar[[1]],PosScalar[[1]]}]
\end{lstlisting}
See sec.~\ref{sec:AnomDim}
for how to obtain anomalous dimensions for vector-bosons and fermions.

\subsection{Integrating out temporal scalars}

Temporal scalars are often heavy compared to the fields driving the phase transition, and can be integrated out~\cite{Kajantie:1995dw,Farakos:1994kx}.
It should be stressed that this is a user-dependent  optional step.
The resulting EFT is said to describe ultrasoft physics.
The command is
\begin{lstlisting}[language=Mathematica,mathescape=true]
PerformDRsoft[{}];
\end{lstlisting}
Couplings at the ultrasoft scale are given by
\begin{lstlisting}[language=Mathematica,mathescape=true]
PrintCouplingsUS[];
\end{lstlisting}
And effective scalar masses are given by
\begin{lstlisting}[language=Mathematica,mathescape=true]
PrintScalarMassUS["LO"];
PrintScalarMassUS["NLO"];
\end{lstlisting}
Finally, $\beta$-functions for the ultrasoft masses are provided with the command
\begin{lstlisting}[language=Mathematica,mathescape=true]
BetaFunctions3DUS[];
\end{lstlisting}
In the 3-dimensional theory only scalar masses have non-zero beta functions.

This concludes our tutorial of
the dimensional reduction of the Abelian-Higgs model.

%
\subsection{Two-loop effective potential}

Once the model is loaded, we can calculate the effective potential.
This can be done by the user, either in the soft or the ultrasoft theory.
Alternatively,
\dralgo{} can calculate the effective potential in the ultrasoft theory. 
First, we need to create field-dependent masses
which requires to specify a \vev{} direction:
\begin{lstlisting}[language=Mathematica,mathescape=true]
DefineNewTensorsUS[$\mu$ij,$\lambda$4,$\lambda$3,gvss,gvvv];
$\phi$vev={0,$\phi$}//SparseArray;
DefineVEVS[$\phi$vev];
\end{lstlisting}
Here the \vev{} is in the second, imaginary, Higgs component.
	
For the calculation to proceed,
all mass matrices must be diagonal.
If not, an error message is printed and it is up for the user
to diagonalize the matrices;
see section~\ref{sec:NonDiagonal} for an example.
For the model at hand,
the mass matrix is diagonal,
which the user can confirm by printing the field-dependent masses
\begin{lstlisting}[language=Mathematica,mathescape=true]
FieldMasses=PrintTensorsVEV[]
\end{lstlisting}
The effective potential is calculated via
\begin{lstlisting}[language=Mathematica,mathescape=true]
CalculatePotentialUS[]
\end{lstlisting}
The results are given by
\begin{lstlisting}[language=Mathematica,mathescape=true]
PrintEffectivePotential["LO"]
PrintEffectivePotential["NLO"]
PrintEffectivePotential["NNLO"]
\end{lstlisting}
Here,
LO refers to the tree-level,
NLO to the one-loop, and
NNLO to the two-loop effective potential.
Note that all results are given in Landau gauge.
In the (two-loop) NNLO part, the renormalization scale is denoted by
${\tt \mu3}=\Lamd$.

This completes the first tutorial on quick installation and running.
Output from the 3d EFT matching relations can be implemented either to
non-perturbative lattice codes, or
perturbative analyses in terms of the effective potential.
While such implementations are left to the user,
appendix~\ref{sec:thermo} displays how to interface \dralgo{} output in
a {\tt Mathematica} implementation that determines
selected thermodynamic quantities for the Abelian-Higgs model.

%
\section{Theory behind the scenes: dimensional reduction of a generic model}
\label{sec:theory}

In this section,
we dig deeper into the theory and computations in the back-end of the software.
The implementation for dimensional reduction of a generic model is based on
tensor-notation~\cite{%
  Martin:2017lqn,Martin:2018emo,Machacek:1984zw,Machacek:1983fi,Machacek:1983tz}
that separates Lorentz algebra from group algebra.
The Lorentz algebra is hard-coded, with contractions done by
{\tt FeynCalc}~\cite{Shtabovenko:2016sxi,Shtabovenko:2020gxv} and
{\tt FORM}~\cite{Ruijl:2017dtg,Croon:2020cgk}, while
the group-structure is provided by the user.

\subsection{Lagrangian in the 4d fundamental theory}

Consider a general Lagrangian in Minkowski spacetime with
mostly-plus metric: 
$g^{\mu \nu}=\mbox{diag}(-1,1,1,1)$,
$\left\{\sigma_\mu,\ol{\sigma}_\nu \right\}=-2 g_{\mu \nu}$.
In terms of
the functional-integral measure and
the action $S$, the partition function is
$\mathcal{Z}=\int \mathcal{D}\,e^{i S}$.
Correspondingly,
the sigma matrices are defined as
\begin{align}
  \sigma^{\mu}=\left(\mathbbm{1},\sigma^i\right),\quad
  \overline{\sigma}^{\mu}=\left(-\mathbbm{1},\sigma^i\right)
  \;,
\end{align}
where $\sigma^i$ are Pauli matrices.

To write down the Lagrangian, in a flattened form, we employ a basis where
all scalars and vectors are real.
In addition, all fermions are composed of
two-component Weyl-spinors~\cite{Dreiner:2008tw}.
The most general, four dimensional, Lagrangian in
Minkowski space is~\cite{Martin:2017lqn,Martin:2018emo,Braaten:1995cm}
\begin{align}
\mathcal{L}&=
  - \frac{1}{2} R_{i}^{ }(-\delta_{ij}^{ }\partial_\mu^{ } \partial^\mu + \mu_{ij}) R_{j}^{ }
  - \frac{1}{4}F_{\mu \nu}^a F^{\mu \nu,a}\delta_{ab}
  - \frac{1}{2 \xi_a}(\partial_\mu A^{a,\mu})^2
  \nn &
  - \partial^\mu \overline{\eta}^a \partial_\mu \eta^a
  + i\psi_{I}^{\dagger}\overline{\sigma}^\mu \partial_\mu \psi^{I}
  - \frac{1}{2}(M^{IJ}\psi_{I}^{ } \psi_{J}^{ } + \text{h.c.})
  + \mathcal{L}_{\rmi{int}}
  \;,
  \\[3mm]
\mathcal{L}_{\rmi{int}}&=
 -\lambda_i^{} R_i^{} - \frac{1}{3!}\lambda_{i j k}^{ }R_{i}^{ } R_{j}^{ } R_{k}^{ }
  - \frac{1}{4!}\lambda_{jkl m}^{ }R_{i}^{ } R_{j}^{ } R_{k}^{ } R_{m}^{ }
  - \frac{1}{2}(Y^{i I J }R_{i}^{ } \psi_{I}^{ } \psi_{J}^{ } + h.c)
  \nn &
  + g_{IJ}^{a}A^a_\mu \psi_{I}^{\dagger}\overline{\sigma}^\mu \psi_{J}^{ }
  - g_{jk}^{a}A^a_\mu  R_{j}^{ } \partial^\mu R_{k}^{ } 
  - \frac{1}{2}g^{a}_{j n}g^b_{kn}A_\mu^a A^{\mu,b}R_{j}^{ } R_{k}^{ }
  - g^{abc}A^{\mu,a}A^{\nu,b}\partial_\mu^{ } A^c_\nu
  \nn &
  - \frac{1}{4}g^{abe}g^{cde}A^{\mu a}A^{\nu b}A_{\mu}^c A_\nu^d
  + g^{abc}A_\mu^a \eta^b \partial^\mu \overline{\eta}^c
  \;,
\end{align}
where
$\mathcal{L}_{\rmi{int}}$ is the interaction Lagrangian and
$F_{\mu \nu} = \frac{i}{g}[D_\mu,D_\nu]$ is the field strength tensor
with the corresponding gauge coupling $g$.
In this notation,
the field $R_i$ denotes a real scalar-field with scalar-index $i$.
In the Standard Model, $i$ corresponds to
the Higgs,
neutral Goldstone, and
real/imaginary component of the charged Goldstone.
Further,
the field $A_\mu^a$ corresponds to a real vector-field with vector-index $a$,
the field $\eta$ is a ghost-field and
the field $\psi_I$ is a Weyl-spinor with fermion-index $I$.
In addition,
$\mu_{ij}$ denote (squared) scalar masses, and
$M_{IJ}$ fermion masses.
Repeated indices are always summed over irrespective of their vertical placement.

Above we denoted
$Y_{i I J}$ as Yukawa couplings,
$\lambda_{i j k l}$ as scalar quartic,
$\lambda_{i j k}$ as scalar cubic, and 
$\lambda_{i }$ as scalar tadpole couplings.
The gauge couplings
$g^{a}_{j k}$ are (anti-symmetric) representation matrices.
For example,
for an adjoint scalar-representation in $\mathrm{SO}(3)$
$g^{a}_{j k}=\epsilon^{a j k}$. 
The abbreviation h.c.\ stands for the hermitian conjugate.
The vertical placement of fermion indices is important;
e.g.\ $M^{IJ}=M^{*}_{IJ}$.
We refer to~\cite{Dreiner:2008tw} for further details.

In the dimensional-reduction step, hard modes with masses $\sim \pi T$
are integrated out.
This is done by matching
the fundamental Lagrangian above to
the effective Lagrangian living in three-dimensions. 
To avoid large logarithms this matching should be performed close to
$\bmu=\pi T$
where $\bmu$ is the RG-scale.
This three-dimensional Lagrangian does not contain fermions.
Moreover, in the dimensional-reduction step the temporal component of vectors -- represented by temporal scalar fields in the EFT -- obtain Debye masses, as well as thermally generated interactions with other scalars.

The matching is straightforward using Euclidean signature.
This is achieved by redefining
\begin{align}
  A^{\rmii{M},a}_0 &\equiv i A^{\rmii{E},a}_0
  \;, \quad
  \sigma_i^{\rmii{M}}=-i \sigma_i^{\rmii{E}}
  \;, \quad
  t=-i \tau
  \;, \quad
  \partial_0^{\rmii{M}}=i \partial_0^{\rmii{E}}
  \;,
  \\[2mm]
  \left(\partial_\mu R \,\partial^\mu R\right)_{\rmii{M}} &=
  \left(\partial_\mu R \, \partial^\mu R\right)_{\rmii{E}}
  \;, \quad
  \bsl{D}_{\rmii{M}}=i \bsl{D}_{\rmii{E}}
  \;, \quad
  \left\{\sigma_\mu,\ol{\sigma}_\nu \right\}_{\rmii{E}} =
  2 \delta_{\mu \nu}
  \;,
\end{align}
where quantities are denoted either in
Minkowskian ($M$) or
Euclidean ($E$) metric. Henceforth,
we suppress these subscripts and implicitly assume an Euclidean metric.

With the above redefinitions, the partition function is defined as
$\mathcal{Z}_{\rmii{E}} =\int\mathcal{D}\,e^{-S_{\rmii{E}}}$ with
the most general tree-level Lagrangian in
Euclidean signature
\begin{align}
\mathcal{L}&=
   \frac{1}{2} R_{i}^{ }(\delta_{ij}^{ }\partial_\mu^{ }\partial^\mu + \mu_{ij}) R_{j}^{ }
  + \frac{1}{4}F_{\mu \nu}^a F^{\mu \nu,a}\delta_{ab}
  + \frac{1}{2\xi_a}(\partial_\mu A^{a,\mu})^2
  \nn &
  + \partial^\mu \overline{\eta}^a \partial_\mu \eta^a
  + \psi_{I}^{\dagger}\overline{\sigma}^\mu \mathcal{\partial}_\mu \psi^{I}
  + \frac{1}{2}(M^{IJ}\psi_{I}^{ }\psi_{J}^{ }+\mathrm{h.c.})
  + \mathcal{L}_{\rmi{int}}
  \;,\\[3mm]
\mathcal{L}_{\rmi{int}}&=
    \lambda_{i}^{ }R_{i}^{ }+ \frac{1}{3!}\lambda_{i j k}^{ }R_{i}^{ } R_{j}^{ } R_{k}^{ }
  + \frac{1}{4!}\lambda_{jklm}^{ } R_{i}^{ } R_{j}^{ } R_{k}^{ } R_{m}^{ }
  + \frac{1}{2}(Y^{i I J}R_{i}^{ } \psi_{I}^{ } \psi_{J}^{ } + \mathrm{h.c.})
  \nn &
  + ig_{IJ}^{a}A^a_\mu \psi_{I}^{\dagger}\overline{\sigma}^\mu \psi_{J}^{ }
  + g_{jk}^{a} A^a_\mu  R_{j}^{ } \partial^\mu R_{k}^{ }
  + \frac{1}{2}g^{a}_{j n}g^b_{kn}A_\mu^a A^{\mu,b}R_{j}^{ } R_{k}^{ }
  + g^{abc}A^{\mu,a}A^{\nu,b}\partial_\mu A^c_\nu
  \nn &
  + \frac{1}{4}g^{abe}g^{cde}A^{\mu a}A^{\nu b}A_{\mu}^c A_\nu^d
  - g^{abc}A_\mu^a \eta^b \partial^\mu \overline{\eta}^c
  \;.
\end{align}
All Lorentz indices are contracted using the
Euclidean metric 
$\delta_{\mu\nu}=\mbox{diag}(+1,+1,+1,+1)$.
The Feynman rules for the vertices are
\begin{align}
  R_{i}R_{j}R_{k}R_{l} &:
  -\lambda_{ijkl}
  \;,\\
  R_{i}R_{j}R_{k} &:
  -\lambda_{ijk}
  \;,\\
    R_{i} &:
  -\lambda_{i}
  \;,\\
  R_{i}\psi_{I}\psi_{J} &:
  -Y_{i IJ}
  \;,\\
  \psi_{I}^{ }\psi_{J}^{ }A^a_\mu &:
  -ig^{a}_{IJ} \overline{\sigma}_\mu
  \;,\\
  R_{i}^{ }R_{j}^{ } A^a_\mu &:
  i g_{ij}^a (p+q)_\mu
  \;,\\
  A^a_\mu A^b_\nu A^c_\rho &:
  i g^{abc}T_{\mu\nu\rho}
  \;,\\
  A^a_\mu A^b_\nu A^c_\rho A_\sigma^d &:
  -G_{\mu\nu\rho\sigma}^{abcd}
  \;,\\
  \overline{\eta}^a \eta^b A^c_\mu &:
  ig^{abc}p_\mu
  \;,\\
  \label{eq:RRAA}
  R_{i}^{ }R_{j}^{ } A^b_\mu A^b_\nu &:
 \delta _{\mu \nu}^{ }\bigl(g_{in}^a g_{n j}^b+g_{jn}^a g_{n i}^b\bigr)\equiv \delta_{\mu \nu}H^{ab}_{ij}
  \;,
\end{align}
where
$G_{\mu\nu\rho\sigma}^{abcd}$ and
$T_{\mu\nu\rho}$ are defined in~\cite{Martin:2018emo}.
Since we only focus on the matching, only the symmetric phase is relevant and
\vev{}s are not introduced. 
The propagators are
\begin{align}
\label{eq:prop:4d}
  \langle R_i (p) R_j (q) \rangle &=
  \frac{\delta_{ij}\delta(p+q)}{p^2}
  \;, \nn
  \langle A^a_\mu(p) A^b_\nu(q) \rangle &=
 \frac{\delta^{ab}\delta(p+q)}{p^2}  P_{\mu \nu}(p)
  \;, \nn
  \langle   \overline{\eta}^a(p) \eta^b(q) \rangle &=
  \frac{\delta^{ab}\delta(p-q)}{p^2}
  \;, \nn
  \langle \psi_{I}(p) \psi_{J}(q) \rangle &=
  \frac{\delta_{I J}\delta(p-q)i p_\mu \sigma_\mu}{p^2}
  \;,
\end{align}
where all momenta are incoming by convention and
\begin{align}
\label{eq:prop:gauge}
 P_{\mu\nu}(p)=  \delta_{\mu\nu} - (1-\xi)\frac{p_\mu p_\nu}{p^2}
  \;,
\end{align}
where $\xi$ is the corresponding gauge parameter which for our investigations in
Landau gauge is set to $\xi = 0$.
All the Lorentz structure is contained in the vertices, and
the generalized coupling constants take care of the group structure. 

To do the matching we need to renormalize our theory.
The matching can, and  will, introduce kinetic mixing.
To allow for this, we express the bare $(b)$ fields and scalar masses as
\begin{align}
  R_{i(b)}=Z^{1/2}_{ij} R_j
  \;,\quad
  \mu_{ij(b)}=Z^\mu_{ij}=\mu_{ij}^{ }+\delta\mu_{ij}^{ }
  \;,\quad
  {A}^{a}_{\mu(b)}=Z^{1/2}_{ab} A^{b}_{\mu}
  \;,\quad
  \psi_{I(b)}=Z^{1/2}_{IJ}\psi_{J}^{ }
  \;.
\end{align}
By construction all propagators should be diagonal at tree-level, to wit
\begin{align}
  Z^{1/2}_{i k}Z^{1/2}_{k j}=\delta_{ij}+\delta Z_{ij}
  \;,\quad
  Z^{1/2}_{a c}Z^{1/2}_{c b}=\delta^{ab}+\delta Z^{ab}
\;.
\end{align}

\subsubsection*{Assumed size of masses and couplings}

For the current problem there are three energy scales:
$\pi T$, $g T$, and $g^2 T$.
Denoted as the hard, soft, and ultrasoft scale, respectively.
In \dralgo{} it is assumed that all scalar (fermion) masses scale as $g T$ or $g^2 T$.
If, on the contrary, some masses would be hard, then they should formally be integrated out together with high-temperature modes.%
\footnote{
  This will be implemented in future versions of \dralgo{}.
}
After hard modes are integrated out, it is up to the user whether soft-scale ($g T$) fields are integrated out as well.
By default all temporal scalars have masses of this order,
but some models might also have additional scalars with soft masses.
For the couplings, \dralgo{} assumes that all quartics and cubics scale as $g^2$. This scaling is chosen so that scalar masses of $\mathcal{O}(g T)$ can be
integrated out safely~\cite{Niemi:2021qvp}.

In summary, \dralgo{} assumes the following scaling
\begin{align}
\label{eq:AssumedScaling}
  \lambda_{ijkl} &\sim g^2\;,
  &
  \lambda_{ijk} &\sim g^2\;,
  &
  \lambda_{i} &\sim g^2\;,
  \nn
  \mu_{ij} &\sim g^2 T^2\;,
  &
  M_{IJ} &\sim g T\;,
  &
  Y_{i IJ} &\sim g\;,
  &
  g^{a}_{IJ} &\sim g^{a}_{ij}\sim g^{abc}\sim g
  \;.
\end{align}
It is of course possible to consider smaller couplings and masses than those above.

\subsection{Lagrangian in the 3d effective theory}

The three-dimensional theory is of the form
\begin{align}
\mathcal{L}&=
    \frac{1}{2} R_{i}^{ }(\delta_{ij}^{ }\partial_\mu^{ } \partial^\mu+\mu_{ij}) R_{j}^{ }
  + \frac{1}{4}F_{\mu\nu}^a F^{\mu\nu,a}\delta_{ab}
  + \frac{1}{2\xi_a}(\partial_\mu A^{a,\mu})^2
  \nn &
  + \partial^\mu \overline{\eta}^a \partial_\mu \eta^a
  + \frac{1}{2} \partial_\mu A_0^a \partial^\mu A_0^a
  + \frac{1}{2} \mu_{\rmii{D}}^{ab} A_0^a  A_0^b
  + \mathcal{L}_\text{int}
  \;,\\
\mathcal{L}_\text{int}&=
   h_{i} R_{i}+ \frac{1}{3!}h_{ijk}^{ } R_{i}^{ } R_{j}^{ } R_{k}^{ }
  + \frac{1}{4!}h_{jklm}^{ }R_{i}^{ } R_{j}^{ } R_{k}^{ } R_{m}^{ }
  \nn &
  + g_{jk}^{a} A^a_\mu  R_{j}^{ } \partial^\mu R_{k}^{ }
  + \frac{1}{2}A_\mu^a A^{\mu,b}R_{j}^{ } R_{k}^{ }
  + g^{abc}A^{\mu,a}A^{\nu,b}\partial_\mu A^c_\nu
  \nn &
  + \frac{1}{4}g^{abe}g^{cde}A^{\mu a}A^{\nu b}A_{\mu}^c A_\nu^d
  - g^{abc}A_\mu^a \eta^b \partial^\mu \overline{\eta}^c
  \nn &
  + \lambda_A^{abcd}A_0^a A_0^b A_0^c A_0^d
  + {g_0}_{ij}^{ab} A_0^a A_0^b R_{i}^{ } R_{j}^{ }
  + g_{0}^{abc} A_\mu^a A_0^b \partial^\mu A_0^c
  \;,
\end{align}
where Lorentz indices range between $\mu,\nu =\{1,\dots,3\}$.
In renormalizing the fields,
we define 
\begin{align}
  Z^{1/2}_{i k}Z^{1/2}_{k j}=
    \delta_{ij}^{ }
  + \delta Z_{3,ij}
  \;,\quad
  Z^{1/2}_{a c}Z^{1/2}_{c b} = 
    \delta^{ab}
  + \delta Z^{ab}_3
  \;,\quad
    Z^{1/2}_{a c,\rmii{L}}
    Z^{1/2}_{c b,\rmii{L}} =
    \delta^{ab}
  + \delta Z^{ab,\rmii{L}}_3
  \;.
\end{align}
Here,
the term $\delta Z^{ab,\rmii{L}}_3$ corresponds to temporal vectors.
The propagators of this theory are
as in eq.~\eqref{eq:prop:4d} in three dimensions with
the absence of the fermionic propagator and
the inclusion of the temporal-vector propagator
\begin{align}
\label{eq:prop:3d}
  \langle A_0^a (p) A_0^b (q) \rangle &= \frac{\delta^{ab}\delta(p+q)}{p^2}
  \;.
\end{align}
The counterterm Feynman rules are the same as above,
barring the new temporal-vector diagram
\begin{align}
A_0^a A_0^b:
  - p^2\delta Z^{ab,\rmii{L}}_3
  - \delta \mu_{\rmii{D}}^{ab}.
\end{align}

%

\subsection{Matching}

Self-energies and
general $n$-point correlations are calculated both in
the original 4d theory and
the effective 3d theory.
The 3-dimensional parameters are chosen, or matched, to give the same correlators as the original theory~\cite{Farakos:1994kx,Kajantie:1995dw}.

\subsubsection*{One-loop thermal scalar masses}
We start with the matching for scalar masses.
\begin{figure}[t]
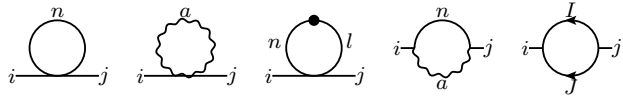

\begin{equation*}
  \TopoSTtxt(\Lsai,\Asai,i,j,n)\quad
  \TopoSTtxt(\Lsai,\Agli,i,j,a)\quad
  \TopoSTxtxt(\Lsai,\Asai,i,j,n,l)\quad
  \TopoSBtxt(\Lsai,\Asai,\Agli,i,j,n,a)\quad
  \TopoSBtxt(\Lsai,\Aqu,\Auq,i,j,I,J)
\end{equation*}
\caption{%
  Diagrams contributing to the thermal scalar masses.
  A wiggly line denotes gauge bosons,
  a black line scalar particles, and
  a directed black line fermions. 
  A black dot represents corresponding counterterms.
}
\label{fig:2pt}
\end{figure}
For the matching we assume to be in a regime 
of soft or ultrasoft external momenta
$p\sim gT$ or
$p\sim g^2T$.
The corresponding diagrams at one-loop level are illustrated
in fig.~\ref{fig:2pt}.
In their computation,
we need the master integrals~\cite{Braaten:1995jr}
\begin{align}
\label{eq:master:1l:b}
  I^{4b}_{\alpha} &= \Tint{Q}\frac{1}{[Q^2]^\alpha}=
  \left(\frac{\bmu^2e^\gammaE}{4\pi}\right)^\epsilon
  2T
  \frac{[2\pi T]^{d-2\alpha}}{(4\pi)^{\frac{d}{2}}}
  \frac{\Gamma\left(\alpha - \frac{d}{2}\right)}{\Gamma(\alpha)}
  \zeta_{2\alpha-d}
  \;,\\
\label{eq:master:1l:f}
  I^{4f}_{\alpha} &= \Tint{\{Q\}}\frac{1}{[Q^2]^\alpha}=
  \left(\frac{\bmu^2e^\gammaE}{4\pi}\right)^\epsilon
  2T
  \frac{[2\pi T]^{d-2\alpha}}{(4\pi)^{\frac{d}{2}}}
  \frac{\Gamma\left(\alpha - \frac{d}{2}\right)}{\Gamma(\alpha)}
  \bigl(1-2^{d-2\alpha}\bigr)\zeta_{2\alpha-d}
  \;,
\end{align}
where $\zeta_s = \zeta(s)$ is the Riemann zeta function.
The $d$-dimensional bosonic integral measure is defined as
\begin{align}
\Tint{Q} \equiv
T\sum_{q_n}  \left(\frac{\bmu^2e^\gammaE}{4\pi}\right)^\epsilon
\int\frac{{\rm d}^{d}\vec{q}}{(2\pi)^d}
\;,	
\end{align}
while
for fermions the summation is written as
$\Tinti{\{Q\}}$.
We set $d=3-2\epsilon$
with Euclidean four-momenta
$Q^2 = q_n^2 + \vec{q}^2 =
    \bigl[ (2n + \sigma)\pi T \bigr]^2
  + \vec{q}^2
$, and
$\sigma = 0(1)$
for bosons(fermions).

The first diagram in fig.~\ref{fig:2pt} gives
\begin{align}
  D_1^{ij} &=-\frac{1}{2}\lambda_{i j n n} I^{4b}_{1}
  \;, 
\end{align}
where the factor of $\frac{1}{2}$ is the symmetry factor,
$-\lambda_{ijnn}$ is the Feynman rule, and 
\begin{align}
  I^{4b}_{1} &=
      \frac{T^2}{12} 
    + \mathcal{O}(\epsilon)
  \;.
\end{align}
The zero Matsubara mode is here ignored.
The reason for this is that the zero Matsubara mode corresponds to soft momenta, and should not be integrated out.
Conversely,
since the $n=0$ mode exists both in
the EFT and the parent theory, this contribution cancels in the matching.
The second diagram is likewise (neglecting higher $\epsilon$ terms)
\begin{align}
  D_2^{ij} &= \frac{1}{2}H^{aa}_{ij} dI_1^{4b}
  \;,\quad
dI_1^{4b} = \frac{T^2}{4}
  \;,
\end{align}
where we used the shorthand notation for
$H^{a b}_{i j}$ from eq.~\eqref{eq:RRAA}.

The remaining diagrams give
\begin{align}
  D_3^{ij} =
  i^2 g^a_{i n}g^a_{n j} \frac{3}{(4\pi)^2 \epsilon_b}p^2
  \;, \quad
  D_4^{ij} =
  \frac{1}{2}\lambda_{i j n l}^{ }\, \mu_{n l}^{ } \frac{1}{(4\pi)^2\epsilon_b}
  \;.
\end{align}
The first diagram contains the external momentum is $p$, and
the last diagram is the contribution from a scalar-mass insertion.
With the assumed power-counting~\eqref{eq:AssumedScaling},
this contribution is formally of higher order.
Finally,
there is the fermion loop
\begin{align}
  D_5^{ij} &= \frac{1}{2}\bigl(
    Y^{i I J}Y_{j I J}
  + Y_{i I J}Y^{j I J}\bigr)
  \biggl[ -\frac{T^2}{12}-p^2\frac{1}{(4\pi)^2\epsilon_f}\biggr]
  \;.
\end{align}
The second coupling-constant term above is the hermitian conjugate.
Also, above we use
$\epsilon_b$ and $\epsilon_f$ to denote $\epsilon$ poles, albeit with some additional factors. They are defined
in~\cite{Kajantie:1995dw,Farakos:1994kx}, and are
\begin{align}
\frac{1}{\epsilon_b}=\frac{1}{\epsilon}+\text{\tt Lb},\quad  \frac{1}{\epsilon_f}=\frac{1}{\epsilon}+\text{\tt Lf},
\end{align}
where
{\tt Lb} and
{\tt Lf} are given in eq.~\eqref{eq:lblf}.

To perform the matching,
we demand
\begin{align}
\int {\rm d}^4x 
\langle R_i R_j\rangle_\rmii{4d}=
T^{-1}\int {\rm d}^3x
\langle R_i R_j\rangle_\rmii{4d}=
\int {\rm d}^3x 
\langle R_i R_j\rangle_\rmii{3d}
\;.
\end{align}
At leading order, 3d fields are rescaled by a factor $T^{-1/2}$.
Next, consider the self-energies.
One finds 
\begin{align}
\bigl(
  - \delta Z_{i j}p^2
  - \delta \mu_{i j}
  - \mu_{i j}
  + \Pi(0)_{i j}
  + p^2 \Pi'(0)_{i j}
  \bigr)_\rmii{4d} =
\bigl(
  - \delta Z_{3,ij} p^2
  - \delta \mu_{3,i j}
  - \mu_{3,i j}
  \bigr)_\rmii{3d}
  \;,
\end{align}
where $\Pi_{ij}(p)=(D_1+\dots+D_5)_{ij}$.
To avoid confusion we added a subscript $3$ to three-dimensional quantities.

Here
$-\delta Z_{ij}p^2-\delta \mu_{i j}$
are $4$d counterterms, and
cancel all $\epsilon$ poles.
The matching gives
\begin{align}
  \mu_{3,i j}=\mu_{i j}-\Pi(0)_{i j}
  \;,\quad 
  \delta Z_{3,i j}=\delta Z_{i j}- \Pi'(0)_{i j}
  \;.
\end{align}
In this step all the Lorentz structure is stripped away,
leaving mere group-theory factors.

%
\subsubsection*{One-loop scalar quartics}
Next, we consider scalar quartics.
Before calculating the actual diagrams, it is 
useful to look at what happens on the 3d side.
The relevant terms in the Lagrangian are
\begin{align}
R_i R_j R_k R_l\,h_{i j k l}
  \;.
\end{align}
There are two contributions to the matching.
First, the coupling constant, $h$,
receives contributions from different loop orders
\begin{align}
h_{i j k l}^{ }=
    h^0_{i j k l}
  + h^1_{i j k l}
  + \ldots
\end{align}
Second, as shown in the previous section, $3$d fields are renormalized.
In terms of the scalar counterterm, $\delta Z_{3,i j}$,
this renormalization is (to leading order)
$ R_{i(b)}=R_i+\frac{1}{2}\delta Z_{3,i j}R_j$.
Combined, these considerations imply that the contribution from the 3d side is
\begin{align}
  \langle R_i R_j R_k R_l \rangle_\rmii{3d} &=
  - h_{i j k l}^0
  - h_{i j k l}^1
  \nn &
  - \frac{1}{2}\left(
    \delta Z_{3,i m}^{ }h_{m j k l}^0
  + \delta Z_{3,j m}^{ }h_{i m k l}^0
  + \delta Z_{3,k m}^{ }h_{i j m l}^0
  + \delta Z_{3,l m}^{ }h_{i j k m}^0
  \right)
  \;.
\end{align}

\begin{figure}[t]
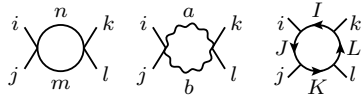

\begin{equation*}
  \TopoVBlrtxt(fex(\Lsai,\Lsai,\Lsai,\Lsai),\Asai,\Asai,ftxt(i,j,k,l,n,m))\quad
  \TopoVBlrtxt(fex(\Lsai,\Lsai,\Lsai,\Lsai),\Agli,\Agli,ftxt(i,j,k,l,a,b))\quad
  \TopoVDtxt(fex(\Lsai,\Lsai,\Lsai,\Lsai),\Aqu,\Aqu,\Aqu,\Aqu,ftxt(i,j,k,l,I,J,K,L))
\end{equation*}
\caption{%
  Diagrams contributing to the thermal corrections to the scalar quartics.
  Indices correspond to the line notation in fig.~\ref{fig:2pt}.
}
\label{fig:4pt}
\end{figure}
In the $4$d theory,
the first and second diagrams give
\begin{align}
  D_1^{ijkl} &= \frac{1}{2}\Bigl(
      \lambda_{i j n m}\lambda_{n m k l}
    + \lambda_{i k n m}\lambda_{n m j l}
    + \lambda_{i l n m}\lambda_{n m k j}
  \Bigr) \frac{1}{16\pi^2 \epsilon_b}
  \;,\\
  D_2^{ijkl} &= \frac{1}{2}\Bigl(
    H^{a b}_{i j}H^{a b}_{k l}
  + H^{a b}_{i k}H^{a b}_{j l}
  + H^{a b}_{i l}H^{a b}_{k j}
  \Bigr)\biggl[ \frac{3}{(4\pi)^2 \epsilon_b}-\frac{1}{8 \pi}\biggr]
  \;.
\end{align}
Finally, the fermion contribution is 
\begin{align}
  D_3^{ijkl} &= \frac{1}{2}\bigl(
    Y_{i I J}Y^{jJ K}Y_{k K L}Y^{l L I}
  + (jkl)
  \bigr)
  \frac{1}{(4\pi)^2 \epsilon_f} 
  + \mathrm{h.c.}
\end{align}
where
the terms $(jkl)$ contain all permutations over indices $j,k,l$.

The matching is performed by demanding
\begin{align}
  \int {\rm d}^4 x
  \langle R_i R_j R_k R_l\rangle_{\rmii{4d}}=
  T^{-1}\int {\rm d}^3 x
  \langle R_i R_j R_k R_l\rangle_{\rmii{4d}}=
  \int {\rm d}^3 x
  \langle R_i R_j R_k R_l\rangle_{\rmii{3d}}
  \;.
\end{align}
In the light of the discussion at the beginning of this section,
this implies
\begin{align}
 T\Bigl(
  - \lambda_{i j k l}
  + \Lambda_{i j k l}\Bigr) &= 
  - h_{i j k l}^0
  - h_{i j k l}^1
  \nn &
  - \frac{1}{2}\left(
    \delta Z_{3,i m}^{ }h_{m j k l}^0
  + \delta Z_{3,j m}^{ }h_{i m k l}^0
  + \delta Z_{3,k m}^{ }h_{i j m l}^0
  + \delta Z_{3,l m}^{ }h_{i j k m}^0
  \right)
  \;.
\end{align}
The renormalized sum of the diagrams above was denoted as
$\Lambda_{i j k l} = (D_1 + D_2 + D_3)_{ijkl}$ 
and
one finds
\begin{align}
  h_{i j k l}^0 &=
  T \lambda_{i j k l},
  \\
  h_{i j k l}^1 T^{-1} &=
  - \frac{1}{2}\Bigl(
      \delta Z_{3,i m}^{ } \lambda_{m j k l}^{ }
    + \delta Z_{3,j m}^{ } \lambda_{i m k l}^{ }
    + \delta Z_{3,k m}^{ } \lambda_{i j m l}^{ }
    + \delta Z_{3,l m}^{ } \lambda_{i j k m}^{ }
    \Bigr)
    - \Lambda_{ i j k l}
    \;,
\end{align}
where
$\delta Z_{3,k m}$ was given in the previous section.

\subsubsection*{Higher loop levels}

When performing dimensional reduction at NLO,
two-loop contributions are required for scalar thermal masses.
Here, we do not show details of
the two-loop computation, as they are analogous to
the one-loop computation presented above.
Naturally the number of required diagrams is larger.
Due to the use of
integration-by-parts identities (IBP) ~\cite{%
  Chetyrkin:1981qh,Tkachov:1981wb,Nishimura:2012ee}
it can be shown that to the given order,
the two-loop master sum-integrals factorize into
the one-loop master integrals given in
eqs.~\eqref{eq:master:1l:b} and \eqref{eq:master:1l:f}.
While this streamlines the computation at NLO significantly,
contributions at higher orders are non-factorizable.
Nonetheless, \dralgo{} also implements
a generic computation of two-loop thermal masses and other 3d parameters
at NLO.

%
\subsection{Beta functions and anomalous dimensions}
\label{sec:BetaFunc}

\dralgo{} calculates beta functions and anomalous dimensions by renormalizing 4d parameters (cf.~\cite{Machacek:1983tz,Machacek:1983fi,Machacek:1984zw}). 
As an example, consider the scalar sector.
We have renormalized our fields and couplings as
\begin{align}
  R_{i(b)}=Z^{1/2}_{ij} R_{j} &= \Bigl(\delta_{ij}+\frac{1}{2}\delta Z_{ij}\Bigr)R_j
  \;, \\
  R_{i(b)} R_{j(b)} R_{k(b)} R_{l(b)}\lambda_{ijkl(b)} &=
  R_i R_j R_k R_l\bigl(\lambda_{ijkl}+\delta\lambda_{ijkl}\bigr)\bmu^{2\eps}.
  \;
\end{align}
On the 4d side,
we assume the $\overline{\text{MS}}$ scheme, so
$\delta Z_{ij}$,
$\delta\mu_{ij},$ and
$\delta\lambda_{ijkl}$ are chosen to {\em only} cancel
$\epsilon$ poles.

We demand that bare parameters are independent of the RG-scale $\bmu$.
To leading order in $\epsilon$, we find
($t = \log \bmu$)
\begin{align}
  \partial_{t}\lambda_{ijkl} =
  - 2\epsilon \lambda_{ijkl}
\;, \quad
\partial_{t}\delta Z_{ij}=-2 \epsilon \delta Z_{ij}.
\end{align}
Using this we can find the anomalous dimensions by demanding that
$\partial_{t}^{ } R_{i(b)}=0$:
\begin{align}
\implies \partial_{t}{R}_i=
  \epsilon Z_{ij} R_j\equiv \gamma_{ij} R_j
  \;,
\end{align}
where we have introduced the anomalous dimension $\gamma_{ij}$.
Then, the quartic beta function is determined by
\begin{align}
  0 = 
  \bmu^{-2\epsilon}\partial_{t}\lambda_{ijkl(b)} &=
    2\epsilon\bigl(\lambda_{ijkl} + \delta\lambda_{ijkl}\bigr)
  + \partial_{t}\delta\lambda_{ijkl}
  + \partial_{t}\lambda_{ijkl}
  \;,\nn
  \implies  \partial_{t}\lambda_{ijkl} &= 
  - 2\epsilon \lambda_{ijkl}
  + 2\epsilon \delta \lambda_{ijkl}
  \;.
\end{align}
Where we have used
$\partial_{t}\delta\lambda_{ijkl}=-4\epsilon \delta\lambda_{ijkl}$;
which follows from counting powers of couplings in
$\delta\lambda_{ijkl}$. We identify
$2\epsilon \delta \lambda_{ijkl}$ with the beta function for
$\lambda_{ijkl}$.

This beta function can be written in an equivalent form by writing the counterterm, for $\lambda$, as the sum of a bare term and field renormalization:
\begin{align}
  \delta \lambda_{ijkl}=\bar{\lambda}_{ijkl}+\frac{1}{2}\left(
    \delta Z_{i m}\lambda_{m j k l}
  + \delta Z_{j m}\lambda_{i m k l}
  + \delta Z_{k m}\lambda_{i j m l}
  + \delta Z_{l m}\lambda_{i j k m}
  \right).
\end{align}
Using this one finds
\begin{align}
\partial_{t}{\lambda}_{ijkl}=
  - 2\epsilon \lambda_{ijkl}
  + 2\epsilon \bar{\lambda}_{ijkl}
  - (\gamma \lambda)_{(ijkl)}
  \;,
\end{align}
where the last term denotes all possible contractions between one index of
$\lambda_{ijkl}$ and one index of
$\gamma_{ij}$.
Other beta functions are calculated analogously.

All gauge beta functions in \dralgo{} are defined with couplings squared,
{\em viz.~$\partial_t^{ } g^2$.}
Conversely beta functions for
masses,
scalar couplings, and
Yukawa couplings are defined
as linear in the parameters, i.e.\
$\partial_t \lambda$ and so forth.
These conventions are also shown in the \dralgo{} output.

In the 3-dimensional theory only scalar masses have non-zero beta functions.

%
\section{Implementing beyond the Standard Model theories}
\label{sec:BSM}

In this section,
we demonstrate the use of \dralgo{} for BSM theories based on
the example of the Two-Higgs doublet model (2HDM).

\subsection{Two-Higgs doublet model with fermions}
\label{sec:2HDM}

Consider the 2HDM potential~\cite{Branco:2011iw,Gunion:1989we}
\begin{align}
\label{eq:V-2hdm}
    V(\phi_1,\phi_2) 
&=
    m_1^2 \phi_1^{ } \phi_1^\dagger
  + m_2^2 \phi_2^{ } \phi_2^\dagger
  - (m_{12}^2 \phi_1^{ } \phi_2^\dagger
    + \text{h.c.} )
  + \lambda_1^{ }(\phi_1^{ } \phi_1^\dagger)^2
  + \lambda_2^{ }(\phi_2^{ } \phi_2^\dagger)^2
  \nn &
  + \lambda_3^{ }(\phi_1^{ } \phi_1^\dagger )(\phi_2^{ } \phi_2^\dagger)
  + \lambda_4^{ }(\phi_1^{ } \phi_2^\dagger )(\phi_2^{ } \phi_1^\dagger)
  + \frac{\lambda_5}{2}\left[
      (\phi_1^{ } \phi_2^\dagger)^2
    + (\phi_2^{ } \phi_1^\dagger)^2
  \right]
  \nn &
  +\left\{
    \left[
        \lambda_6^{ }(\phi_1^{ } \phi_1^\dagger)
      + \lambda_7^{ }(\phi_2^{ } \phi_2^\dagger)
    \right]
  (\phi_1^\dagger \phi_2^{ })
  + \mathrm{h.c.}
  \right\}
  \;.
\end{align}
The scalars are $\mathrm{SU}(2)$ doublets with hypercharge $Y_\phi=1$.
This means that the covariant derivative is given by%
\footnote{
  We use the convention that
  $
  (\phi_1^{ } \phi_1^\dagger)=
  (\phi_1^\dagger \phi_1^{ })=\phi_1^{I} \phi_1^{\dagger,I}$,
  where $I=1,2$.
}
\begin{align}
\label{eq:D:2hdm}
  \mathcal{D}_\mu \phi_{1,2}=
    \Bigl(
      \partial_\mu
      - i g_1 \frac{Y_\phi}{2} B_{\mu}
      - i g_2 \frac{\tau^a}{2} A^a_{\mu}
    \Bigr)\phi_{1,2}
    \;,
\end{align}
where $\tau^a$ are Pauli matrices.
We further assume that all Standard-Model fermions are present, and consider Yukawa interactions of the form
\begin{align}
    \mathcal{L}_\rmi{Yuk}=y_{i}^{ } \overline{q}_L \phi^\dagger_i q_R
  + \mathrm{h.c.} 
\end{align}
We define the group structure and scalar representations via
\begin{lstlisting}[language=Mathematica,mathescape=true]
Group={"SU3","SU2","U1"};
RepAdjoint={{1,1},{2},0};
HiggsDoublet1={{{0,0},{1},1/2},"C"};
HiggsDoublet2={{{0,0},{1},1/2},"C"};
RepScalar={HiggsDoublet1,HiggsDoublet2};
CouplingName={g3,g2,g1};
\end{lstlisting}
All representations are specified according to their Dynkin coefficients
(see e.g.~\cite{Georgi:1999wka,Slansky:1981yr,Fonseca:2020vke}).
For example, gauge bosons transforms as $(1,1)$, or an octet, under $\mathrm{SU}(3)$.
Next we need to create the fermions. Let us start by creating a single generation
\begin{lstlisting}[language=Mathematica,mathescape=true]
Rep1={{{1,0},{1}, 1/6},"L"};  (*$q_\rmii{L}$*)
Rep2={{{1,0},{0}, 2/3},"R"};  (*$u_\rmii{R}$*)
Rep3={{{1,0},{0},-1/3},"R"};  (*$d_\rmii{R}$*)
Rep4={{{0,0},{1},-1/2},"L"};  (*$\ell_\rmii{L}$*)
Rep5={{{0,0},{0},-1},"R"};  (*$e_\rmii{R}$*)
RepFermion1Gen={Rep1,Rep2,Rep3,Rep4,Rep5};
\end{lstlisting}
In the definition of a fermion representation
{\verb!{{{1,0},{1},Y/2},"L<R>"}!}
the last argument depends on if the fermion 
left-handed ({\tt L}) or
right-handed ({\tt R}) and
{\tt Y} is the corresponding hypercharge.
For the Standard Model, we have
\begin{align}
\label{eq:hypercharge}
  \Yq = \frac{1}{3}\;,\quad
  \Yu = \frac{4}{3}\;,\quad
  \Yd = -\frac{2}{3}\;,\quad
  \Yl = -1\;,\quad
  \Ye = -2
  \;.
\end{align}
The above notation identifies
{\tt Rep1} as the left-handed quark doublet,
{\tt Rep2} as the right-handed up-quark and so forth.
The extra factor of $\frac{1}{2}$ for hypercharges in the definition of
e.g.\ {\tt Rep1}
arises from the definition of the covariant derivative similar to eq.~\eqref{eq:D:2hdm}.

Additional fermion-families can be added by grouping multiple instances of
{\tt RepFermion1Gen} together:
\begin{lstlisting}[language=Mathematica,mathescape=true]
RepFermion3Gen={RepFermion1Gen,RepFermion1Gen,RepFermion1Gen}//Flatten[#,1]&;
\end{lstlisting}
where the number of generations is chosen to be $\nf = 3$.
When creating Yukawa interactions the user decides which index contains which family.
For example, above we stacked all generations after each other in {\tt RepFermion3Gen}.
Hence,
in {\tt RepFermion3Gen}
indices 1--5 correspond to the first generation,
indices 6--10 to the second, and
indices 11--15 to the last. 
Thus the user could use index
1, 6, or 11 as the top quark. 
An alternative way to add an arbitrary number of fermion families
$\nf$ is given
in sec.~\ref{sec:nf}.

Next, to create the tensors,
we write
\begin{lstlisting}[language=Mathematica,mathescape=true]
{gvvv,gvff,gvss,$\lambda$1,$\lambda$3,$\lambda$4,$\mu$ij,$\mu$IJ,$\mu$IJC,Ysff,YsffC}=
    AllocateTensors[Group,RepAdjoint,CouplingName,RepFermion3Gen,RepScalar];
\end{lstlisting}
The mass terms in the potential in eq.~\eqref{eq:V-2hdm} are
\begin{align}
    V(\phi_1,\phi_2) &\supset
    m_1^2 \phi_1^{ } \phi_1^\dagger
  + m_2^2 \phi_2^{ } \phi_2^\dagger
  - (m_{12}^2 \phi_1^{ } \phi_2^\dagger+ \mathrm{h.c.})
  \;,
\end{align}
and due to the presence of two Higgs doublets,
we need to specify the doublet for each term:
\begin{lstlisting}[language=Mathematica,mathescape=true]
InputInv={{1,1},{True,False}}; (*$\phi_1^{ } \phi_1^\dagger$*)
MassTerm1=CreateInvariant[Group,RepScalar,InputInv][[1]]//Simplify;
InputInv={{2,2},{True,False}}; (*$\phi_2^{ } \phi_2^\dagger$*)
MassTerm2=CreateInvariant[Group,RepScalar,InputInv][[1]]//Simplify;
InputInv={{1,2},{True,False}}; (*$\phi_1^{ } \phi_2^\dagger$*)
MassTerm3=CreateInvariant[Group,RepScalar,InputInv][[1]]//Simplify;
InputInv={{2,1},{True,False}}; (*$\phi_2^{ } \phi_1^\dagger$*)
MassTerm4=CreateInvariant[Group,RepScalar,InputInv][[1]]//Simplify;
\end{lstlisting}
The mass matrix is then generated by
the {\tt GradMass[]} command: 
\begin{lstlisting}[language=Mathematica,mathescape=true]
VMass=(
    +m1*MassTerm1
    +m2*MassTerm2
    -(m12R+I*m12I)(MassTerm3)
    -(m12R-I*m12I)(MassTerm4)
    );
$\mu$ij=GradMass[VMass]//Simplify;
\end{lstlisting}
We allowed $m_{12}^2$ to be complex, with
real part {\tt m12R}, and
imaginary part {\tt m12I}.
For the scalar quartics in eq.~\eqref{eq:V-2hdm}
the corresponding part of the potential is
\begin{align}
    V(\phi_1,\phi_2) &=
    \lambda_1^{ }(\phi_1^{ } \phi_1^\dagger)^2
  + \lambda_2^{ }(\phi_2^{ } \phi_2^\dagger)^2
  + \lambda_3^{ }(\phi_1^{ } \phi_1^\dagger )(\phi_2^{ } \phi_2^\dagger)
  \nn &
  + \lambda_4^{ } (\phi_1^{ } \phi_2^\dagger )(\phi_2^{ } \phi_1^\dagger)
  + \frac{\lambda_5}{2}\bigl[
      (\phi_1^{ } \phi_2^\dagger)^2
    + (\phi_2^{ } \phi_1^\dagger)^2
    \bigr]
  \nn  &
  + \left\{\bigl[
      \lambda_6^{ }(\phi_1^{ } \phi_1^\dagger)
    + \lambda_7^{ }(\phi_2^{ } \phi_2^\dagger)
  \bigr]
  (\phi_1^\dagger \phi_2^{ })+\mathrm{h.c.}\right\}
  \;,
\end{align}
for which
we already created all the building blocks.
Thus, the quartics can be constructed as
\begin{lstlisting}[language=Mathematica,mathescape=true]
QuarticTerm1=MassTerm1^2; (*$(\phi_1^{ } \phi_1^\dagger)^2$*)
QuarticTerm2=MassTerm2^2; (*$(\phi_2^{ } \phi_2^\dagger)^2$*)
QuarticTerm3=MassTerm1*MassTerm2; (*$(\phi_1^{ } \phi_1^\dagger) (\phi_2^{ } \phi_2^\dagger)$*)
QuarticTerm4=MassTerm3*MassTerm4; (*$(\phi_1^{ } \phi_2^\dagger) (\phi_2^{ } \phi_1^\dagger)$*)
QuarticTerm5=(MassTerm3^2+MassTerm4^2); (*$(\phi_1^{ } \phi_2^\dagger)^2+ (\phi_2^{ } \phi_1^\dagger)^2$*)
QuarticTerm6=MassTerm1*MassTerm3+MassTerm1*MassTerm4; (*$(\phi_1^{ } \phi_1^\dagger)\bigl[ (\phi_1^{ } \phi_2^\dagger)+(\phi_2 \phi_1^\dagger)\bigr]$*)
QuarticTerm7=MassTerm2*MassTerm3+MassTerm2*MassTerm4; (*$(\phi_2^{ } \phi_2^\dagger)\bigl[ (\phi_1^{ } \phi_2^\dagger)+(\phi_2^{ } \phi_1^\dagger)\bigr]$*)
\end{lstlisting}
Consequently, the quartic tensor itself is defined as
\begin{lstlisting}[language=Mathematica,mathescape=true]
VQuartic=(
    +$\lambda$1H*QuarticTerm1
    +$\lambda$2H*QuarticTerm2
    +$\lambda$3H*QuarticTerm3
    +$\lambda$4H*QuarticTerm4
    +$\lambda$5H/2*QuarticTerm5
    +$\lambda$6H*QuarticTerm6
    +$\lambda$7H*QuarticTerm7
    );
$\lambda$4=GradQuartic[VQuartic];
\end{lstlisting}
For simplicity, we assumed above that
{\tt $\lambda$5H},
{\tt $\lambda$6H}, and
{\tt $\lambda$7H} are real.
One can allow for complex couplings by adding them, and their conjugates, directly in
{\tt QuarticTerm6} and
{\tt QuarticTerm7}. 
To create a complex coupling
{\tt $\lambda$6H},
we would write
\begin{lstlisting}[language=Mathematica,mathescape=true]
QuarticTerm6=($\lambda$6HR+I*$\lambda$6HI)*MassTerm1*MassTerm3+($\lambda$6HR-I*$\lambda$6HI)*MassTerm1*MassTerm4;
\end{lstlisting}

For Yukawa couplings, we only consider the top-quark coupling. 
If we choose the first family to contain the top-quark, then
a Yukawa term $\sim\phi^\dagger \overline{q}_L u_R^{ }$ would involve
fermion representations number 1 and 2.
When defining a Yukawa interaction, representations should be specified in the order: scalar, first fermion, second fermion.
With this in mind, the Yukawa coupling of the first Higgs doublet is%
\footnote{
  Here we assume that the top quark resides in the first generation.
  This has nothing to do with how the top-quark is usually placed in the third generation.
  Rather, we here place it in the first for simplicity as normally only the top-quark has a sizeable Yukawa coupling.
}
\begin{lstlisting}[language=Mathematica,mathescape=true]
InputInv={{1,1,2},{False,False,True}};  (*$\phi_1^\dagger \overline{q}_L u_R^{ } $*)
YukawaDoublet1=CreateInvariantYukawa[Group,RepScalar,RepFermion3Gen,InputInv][[1]]//Simplify;
\end{lstlisting}
Here,
{\verb!{1,1,2}!} specifies
$$
\text{First Higgs doublet}\times
\text{Left-handed top-quark}\times
\text{Right-handed top-quark}
\,,
$$
and
{\verb!{False,False,True}!}
specifies that the Higgs and left-handed quarks are conjugated.
The coupling to the second Higgs doublet is
\begin{lstlisting}[language=Mathematica,mathescape=true]
InputInv={{2,1,2},{False,False,True}};  (*$\phi_2^\dagger \overline{q}_L u_R^{ } $*)
YukawaDoublet2=CreateInvariantYukawa[Group,RepScalar,RepFermion3Gen,InputInv][[1]]//Simplify;
\end{lstlisting}
The complete Yukawa tensor is
\begin{lstlisting}[language=Mathematica,mathescape=true]
Ysff=-GradYukawa[yt1*YukawaDoublet1+yt2*YukawaDoublet2];
\end{lstlisting}
For the sake of generality,
we assumed that both Higgs doublets couple to the top-quark.
Note that this is not allowed for a realistic model due to flavor-changing-neutral-current constraints.

Finally, assuming that the Yukawa couplings are real, we can define
\begin{lstlisting}[language=Mathematica,mathescape=true]
YsffC=Simplify[Conjugate[Ysff//Normal],Assumptions->{yt1>0,yt2>0}]//SparseArray;
\end{lstlisting}

Above we did not consider Yukawa couplings between different generations but
such couplings can be added. 
To wit, when we defined {\tt RepFermion3Gen} we put the first generation
on indices 1--5; the second
on indices 6--10; and the third
on indices 11--15.
To define a coupling between, e.g.\
the left-handed top quark (assumed to reside in generation 1) and
the right-handed charm quark (assumed to reside in generation 2),
we would write
\begin{lstlisting}[language=Mathematica,mathescape=true]
InputInv={{1,1,7},{False,False,True}}; 
YukawaTopCharm=CreateInvariantYukawa[Group,RepScalar,RepFermion3Gen,InputInv]//Simplify;
\end{lstlisting}
With the model implementation complete,
all dimensional-reduction commands are identical to those of the Abelian-Higgs model.
We refer to~\cite{Losada:1996ju,Andersen:1998br,Gorda:2018hvi}
for earlier results in the literature.

%
\subsubsection*{Integrating out temporal scalars}
For the Abelian-Higgs case in sec.~\ref{sec:speedrun} we only had one scalar field,
whereas with the 2HDM we have two.
When integrating out temporal scalars we have two options.
First, all scalars are light and we have two active, dynamical doublets.
Second, one doublet is heavy and is integrated out when going from
the soft to
the ultrasoft scale.

For the first case,
the corresponding command is
\begin{lstlisting}[language=Mathematica,mathescape=true]
PerformDRsoft[{}];
\end{lstlisting}
For the second case,
we assume that the second doublet is heavy.
It can then be integrated out via
\begin{lstlisting}[language=Mathematica,mathescape=true]
PerformDRsoft[{5,6,7,8}];
\end{lstlisting}
To understand the above syntax,
note that the indices of all representations are given by 
\begin{lstlisting}[language=Mathematica,mathescape=true]
PosScalars=PrintScalarRepPositions[];
\end{lstlisting}
So the second Higgs doublet resides at {\tt PosScalars[[2]]=5;;8}.

There is one complication when integrating out one of the Higgs doublets. Namely, generally the two doublets mix through the $m_{12}^2$ term.
For a complete treatment the mass matrix needs to be diagonalized before the heavy doublet can be integrated out.
This lies beyond \dralgo{} and is an optional step for the user.
Instead,
the code by default assumes that $m_{12}^2$ is small --
of $\mathcal{O}(g^2T)$ in power counting -- and performs 
the diagonalization perturbatively to first order in $m_{12}^2$.

%
\section{Miscellaneous features}

In this section,
we discuss specific features of the software on a case-by-case basis.

\subsection{User-options and features}

\subsubsection*{Lower-order dimensional reduction for speed}
The user has several options to control what is calculated and how the code operates. 
For example, if the model has many degrees of freedom,
the user might wish to save running-time and only calculate one-loop thermal masses and couplings.
This works by specifying {\tt Mode->1} when loading the model
\begin{lstlisting}[language=Mathematica,mathescape=true]
Group={"U1"};
ImportModelDRalgo[Group,gvvv,gvff,gvss,$\lambda$1,$\lambda$3,$\lambda$4,$\mu$ij,$\mu$IJ,$\mu$IJC,Ysff,YsffC,Mode->1];
\end{lstlisting}
The default is
{\tt Mode->2}, in which case everything is calculated to NLO.
And for the most bare-bone/fast option select
{\tt Mode->0}; in this case only one-loop thermal masses are calculated.

%
\subsection{Model-treatment in the code}

\dralgo{} works by factorizing all group and Lorentz algebra.
The Lorentz algebra is hard-coded while the group algebra is supplied by the user.
All particles are indexed by the order they appear.
If the user has an
$
\mathrm{SU}(3)\times
\mathrm{SU}(2)$ model with gauge bosons,
the vectors in
the $\mathrm{SU}(3)$ group have 8 components, while those of
the $\mathrm{SU}(2)$ group have 3.
The code indexes these components by
$a=\bigl(8^\alpha_{\rmii{SU}(3)},3^\beta_{\rmii{SU}(2)}\bigr)$.
Therein,
$\alpha$ runs from 1--8,
$\beta$ from 1--3, and 
$a$ from 1--11.
For example,
the Debye mass tensor $\mu_{\rmii{D}}^{ab}$
stores the information of both groups. 
For most cases this mass tensor is diagonal.

The scalars are treated differently since
\dralgo{} only deals with real scalar components.
For example, a complex scalar $\Phi$ is rewritten
$\Phi=\frac{1}{\sqrt{2}}\left(\phi +i \psi\right)$.
Hence the scalar indices would in this case be $i=(\phi,\psi)$. Further, the scalar-mass matrix is stored as $\mu_{S,ij}$.
For example, a $m^2 \phi^2$ mass term resides at $\mu_{S,11}$.
Consider now a complex representation with $n$ scalar components
labelled by $I,J$.
In a complex basis vector-scalar-scalar trilinear couplings are of the form
$
A_\mu \left(G_{IJ}^{a}\partial^{\mu}\Phi_{I}^{ } \Phi_{J}^{*}
+ G^{a,*}_{IJ}\partial^{\mu}\Phi_{I}^{*} \Phi_{J}^{ }\right)$,
where $G^a_{IJ}$ are representation matrices.
To convert to a real basis each scalar component is split as 
$\Phi_I=\frac{1}{\sqrt{2}}\left(\phi_I + i \psi_I\right)$.
The components are then reordered with the real components first:
$i=(\phi_I,\psi_I)$, so
$i=1,\ldots,2n$. 
If we call the scalars in the real basis $\varphi_i$,
our vector-scalar-scalar trilinear couplings become
\begin{align}
A_\mu^{ } g^a_{ij}\partial^{\mu}\varphi_i^{ } \varphi_j^{ }
\;,\qquad
g_{ij}^a=
\begin{pmatrix}
    \im(G^a_{IJ}) &
    \re(G^a_{IJ}) \\
  - \re(G^a_{IJ}) &
    \im(G^a_{IJ})
\end{pmatrix}
\;.
\end{align}
Thus,
$g^a_{ij}$ is automatically antisymmetric under $i\leftrightarrow j$.

As an example,
consider a $\mathrm{U}(1)$ theory with
a complex scalar.
The vector-scalar-scalar coupling is
$
i  g A_\mu \left(\partial^\mu \Phi \Phi^{*}-\partial^\mu \Phi^{*} \Phi \right)
$.
In a real basis
$\Phi=\frac{1}{\sqrt{2}}\left(\phi+ i \psi\right) $,
this becomes 
\begin{align}
g A_{\mu}\partial^{\mu}\left(\phi,\psi\right)
\begin{pmatrix}
  0 & \frac{1}{2}\\
  -\frac{1}{2} & 0
\end{pmatrix}
\begin{pmatrix}
  \phi\\
  \psi
\end{pmatrix}
\;,\quad
\text{and}
\quad
g_{ij}=g
\begin{pmatrix}
  0 & \frac{1}{2} \\
  - \frac{1}{2} & 0
\end{pmatrix}
\;.
\end{align}

%
\renewcommand{\thesubsubsection}{Q.\arabic{subsubsection}}
\subsection{Frequently asked questions}
\label{sec:faq}

Below,
we discuss various questions and problems the user might have encountered.

%
\subsubsection{How to save and load my model?}
\label{sec:save}

Saving and loading a model is straightforward with \dralgo{} built-in functions.
Once a model is created, and loaded with {\tt ImportModelDRalgo}
(see the {\tt .m} files in the example folder),
it can be saved by writing
\begin{lstlisting}[language=Mathematica,mathescape=true]
SaveModelDRalgo[ModelInfo,"<modelname>.txt"],
\end{lstlisting}
Here,
{\tt ModelInfo} is a string and should contain information about
the authors,
\dralgo{} version, and
citations to relevant articles.
See
{\tt ./examples/ah.m} or
{\tt ./examples/2hdm.m}
for some examples.
To load the model write
\begin{lstlisting}[language=Mathematica,mathescape=true]
{Group,gvvv,gvff,gvss,$\lambda$1,$\lambda$3,$\lambda$4,$\mu$ij,$\mu$IJ,$\mu$IJC,Ysff,YsffC}=LoadModelDRalgo["<modelname>.txt"];
\end{lstlisting}

The loaded model can be your own, or perhaps one of
the provided \dralgo{} model repository.
We also encourage you to make your own models available for the community
which is possible by submitting the model file
via the Issue Tracker on
\url{https://github.com/DR-algo/DRalgo}. 
The model will then be verified and added to the model repository.
When submitting a model,
please refer to a paper or explicitly write out
the Lagrangian in an accompanying notebook.

%
\subsubsection{How do I order semi-simple groups?}
If the user-defined model contains multiple groups, the groups and corresponding representations must be ordered as:
${\rm SO}(n)$,
${\rm SU}(n)$,
${\rm SP}(n)$,
${\rm G}_2$,
${\rm F}_4$
${\rm E}_6$,
${\rm E}_7$,
${\rm E}_8$,
${\rm U}(1)$.
For example, if the user has a model with gauge group
$
\mathrm{G}=\mathrm{SU}(5)\otimes
\mathrm{SO}(10)\otimes
E_6 \otimes
\mathrm{U}(1)$,
the model-input should be defined as
\begin{lstlisting}[language=Mathematica,mathescape=true]
Group={"SO10","SU5","E6","U1"};
\end{lstlisting}
The internal order of groups is not restricted.
Hence, for a group such as
$\mathrm{G}=
\mathrm{SU}(3)\otimes
\mathrm{SU}(2)$,
the following definitions are equivalent 
\begin{lstlisting}[language=Mathematica,mathescape=true]
Group={"SU3","SU2"};
Group={"SU2","SU3"};
\end{lstlisting}

%
\subsubsection{How do I check that my model is anomaly free?}
Once you have defined your model, the anomaly-free condition is that
\begin{lstlisting}[language=Mathematica,mathescape=true]
Table[Tr[(a.b+b.a).c],{a,gvff},{b,gvff},{c,gvff}]
\end{lstlisting}
vanishes identically.
Since this condition is not automatically fulfilled for
general, non-numeric,
$\mathrm{U}(1)$ charges,
it is the responsibility of the user to choose
$\mathrm{U}(1)$ charges such that all anomalies cancel.

%
\subsubsection{How do I calculate anomalous dimensions?}
\label{sec:AnomDim}

First find the position of all representations:
\begin{lstlisting}[language=Mathematica,mathescape=true]
PosScalar=PrintScalarRepPositions[];
PosVector=PrintGaugeRepPositions[];
PosFermion=PrintFermionRepPositions[];
\end{lstlisting}
All anomalous dimensions are then found via
\begin{lstlisting}[language=Mathematica,mathescape=true]
Table[AnomDim4D["S",{i,j}],{i,PosScalar},{j,PosScalar}]
Table[AnomDim4D["V",{i,j}],{i,PosVector},{j,PosVector}]
Table[AnomDim4D["F",{i,j}],{i,PosFermion},{j,PosFermion}]
\end{lstlisting}
For more details regarding anomalous dimensions and beta functions
cf.\ sec.~\ref{sec:BetaFunc}.

%
\subsubsection{Temporal scalar mixing}
In most models the temporal-scalar masses are diagonal.
However, in cases with multiple $\mathrm{U}(1)$ groups there can be mixing.
In such a case,
the code automatically recognizes this.
For example, if the group is 
\begin{lstlisting}[language=Mathematica,mathescape=true]
Group={"U1","U1","U1"};
\end{lstlisting}
the code creates the mixed U(1) masses with
the convention that
{\tt $\mu$U1Mix1} is the mixing between groups 1 and 2, 
{\tt $\mu$U1Mix2} is the mixing between groups 1 and 3, and
{\tt $\mu$U1Mix3} is the mixing between groups 2 and 3.
These masses are displayed with the {\tt PrintDebyeMass} command.
See the {\tt ./examples/SMZp.m} for an example.

%
\subsubsection{What if I miss some couplings?}
By default \dralgo{} assumes that
the user has defined all couplings allowed by symmetry in their 4d theory.
The code runs even though some couplings are missed.
For example, consider
the 2HDM with scalar potential as in eq.~\eqref{eq:V-2hdm} but 
without
$\lambda_6$ and
$\lambda_7$ couplings.
For a
$\phi_1 \leftrightarrow \phi_2$ symmetric model no
$\lambda_6$/$\lambda_7$-type scalar quartic couplings are induced.
However,
if the model breaks this symmetry by e.g.\ a Yukawa sector, then
$\lambda_7$ and
$\lambda_6$ couplings are generated at one-loop.
In this case \dralgo{} would still calculate these induced couplings;
with the crux that only the
$\lambda_1,\ldots,\lambda_5$ couplings are printed by
the {\tt PrintCouplings[]} command,
while the induced
$\lambda_7$ and 
$\lambda_6$ couplings must be found manually via
the {\tt PrintTensorDRalgo[]} command. 
In addition, \dralgo{} automatically alerts the user with a message
if some new couplings are generated at one-loop.
This is a cross-check that no couplings are forgotten.

%
\subsubsection{Can I run \dralgo{} without specifying the representation?}
For general groups no.
However, it is possible for the user to specify arbitrary (non-numeric)
$\mathrm{U}(1)$ charges.
With non-numeric charges the code does not check that various
quartic or
Yukawa terms are allowed.
Hence, it is the responsibility of the user to ensure gauge invariance.

%
\subsubsection{How large representations can I use?}
In principle any group and representations can be used.
In practice the code slows down for huge representations.
The code has been tested on general models with
$\sim$100--200 components in a given representation.
For example an
$\mathrm{SO}(10)$ model with a $120$-dimensional scalar.
Since \dralgo{} was purposely written to deal with any
quartic and Yukawa sector,
these are often the bottlenecks.
If the user wants to omit two-loop contributions,
significantly larger representations are possible.
Further still,
if the user only wants one-loop thermal masses,
almost any model (within reason) can be run in quick order.

%
\subsubsection{Can I include an arbitrary number of fermion generations?}
\label{sec:nf}

Yes, take for example the 2HDM.
As described in sec.~\ref{sec:2HDM},
a single family of SM fermions is defined via
\begin{lstlisting}[language=Mathematica,mathescape=true]
Rep1={{{1,0},{1}, 1/6},"L"};  (*$q_\rmii{L}$*)
Rep2={{{1,0},{0}, 2/3},"R"};  (*$u_\rmii{R}$*)
Rep3={{{1,0},{0},-1/3},"R"};  (*$d_\rmii{R}$*)
Rep4={{{0,0},{1},-1/2},"L"};  (*$\ell_\rmii{L}$*)
Rep5={{{0,0},{0},-1},"R"};  (*$e_\rmii{R}$*)
RepFermion1Gen={Rep1,Rep2,Rep3,Rep4,Rep5};
\end{lstlisting}
Defining Yukawa and scalar couplings proceed as before, and
the model is loaded with
\begin{lstlisting}[language=Mathematica,mathescape=true]
ImportModelDRalgo[Group,gvvv,gvff,gvss,$\lambda$1,$\lambda$3,$\lambda$4,$\mu$ij,$\mu$IJ,$\mu$IJC,Ysff,YsffC,Verbose->False];
\end{lstlisting}
The user can then define $\nf= {\tt nF}$ fermion families by writing%
\footnote{
  This adds {\tt nF} copies of all fermions.
}

\begin{lstlisting}[language=Mathematica,mathescape=true]
PosFermion=PrintFermionRepPositions[];
FermionMat=Table[{nF,i},{i,PosFermion}];
DefineNF[FermionMat]
\end{lstlisting}
These commands should be used before running
\begin{lstlisting}[language=Mathematica,mathescape=true]
PerformDRhard[];
\end{lstlisting}
Adding $\nf$ fermion families in this way does
not add new Yukawa or scalar couplings as 
the extra $(\nf-1)$ families only have gauge interactions.
To define cross-family Yukawa couplings,
one should follow the procedure in sec.~\ref{sec:2HDM}.

%
\subsubsection{How do I compare with the literature?}
Most outputs from \dralgo{} can directly be compared with the literature.
For temporal scalars this requires some care. 
For example,
if the original 4d theory has a $\mathrm{SU}(2)$ triplet,
then two types of $A_0^2 \Sigma^2$ couplings are allowed:
\begin{align}
  \mathcal{L}_{\rmi{3d}} \supset
  \kappa_1 (A_0^a A_0^a)(\Sigma^a\Sigma^a) +
  \kappa_2 (A_0^a \Sigma^a)^2
  \;.
\end{align}
\dralgo{} does not distinguish between these two terms
since its output is stored in the form
$\frac{1}{4!}A_0^a A_0^b \Sigma^c \Sigma^d \lambda_K^{abcd}$.
To compare with the two parametrizations,
the user can rewrite the
$\kappa_1$/$\kappa_2$ basis in tensor form.
To this end,
it is necessary to additionally define the relevant terms of
the effective model and compare to its Lagrangian.
This procedure is analogous to creating the fundamental model where
invariants can be compared using the command
{\tt CompareInvariants[]}.
See the {\tt htm.m} file for a worked example of
the Higgs triplet model (HTM).

The matching of all possible allowed operators in the EFT is automatic
in \dralgo{}.
This way also previously disregarded effective coefficients can be determined.
One example is a
$\mathcal{L}_{\rmi{3d}} \supset \kappa \tr C_{0}^3B_{0}^{ }$
operator in the Standard Model.
Here,
$C_0$ is the gluon (temporal) field, and
$B_0$ is the temporal hypercharge field.
The output from \dralgo{} gives
$\kappa \propto (2\Yq + \Yd + \Yu)$.
See
eq.~\eqref{eq:hypercharge} for the corresponding hypercharges and
the {\tt 2hdm.m} example file for the full expression

%
\subsubsection{What is the functional basis used in the matching?}
\label{sec:basis}

{In addition to the variables
{\tt Lb} and
{\tt Lf} defined in eq.~\eqref{eq:lblf}, two-loop diagrams also contain factors of
\begin{align}
\label{eq:c}
{\tt c} &=
\frac{1}{2} \Bigl(
  \ln \Big( \frac{8\pi}{9} \Big)
  + (\ln\zeta_{2})'
 - 2 \gammaE \Bigr)
  \;,
\end{align}
where 
$\zeta_{s}=\zeta(s)$ is the Riemann zeta function and
$(\ln \zeta_s)'=\zeta'(s)/\zeta(s)$.
By default \dralgo{}  uses the relations
$\ln(2\pi) - (\ln\zeta_{2})' = 1 - \gammaE + (\ln\zeta_{-1})'$
and
$1+(\ln\zeta_{-1})' = 12\ln A$,
where $A$ is the Glaisher-Kinkelin constant.
It is possible to convert the output of \dralgo{} to
the conventions of~\cite{Kajantie:1995dw}
by using the replacement rule
{\tt Log[Glaisher]->-1/12(Lb+2cplus-EulerGamma)},
where
{\tt cplus=(c+Log[3T/$\mu$])}.
This rule is implemented as
{\tt PrintGenericBasis[]} in \dralgo{}.

%
\subsubsection{How do I define scalar cubic operators?}
Scalar cubics are created analogously to quartics.
Consider for example
a $\mathrm{SU}(2)$ theory with
a scalar doublet $\phi$ and
a singlet $S$.
We can then create the cubic operator
$(\phi \phi^\dagger) S$ via
\begin{lstlisting}[language=Mathematica,mathescape=true]
InputInv={{1,1,2},{True,False,True}}; (*$\phi \phi^\dagger  S$*)
CubicTerm1=CreateInvariant[Group,RepScalar,InputInv][[1]]//Simplify;
VCubic=$\lambda$C*CubicTerm1;
\end{lstlisting}
The corresponding tensor is defined with the command
\begin{lstlisting}[language=Mathematica,mathescape=true]
$\lambda$3=GradCubic[VCubic];
\end{lstlisting}
Tadpoles are defined analogously.
See the
{\tt 2xsm.m}
file for a worked example with two real scalars.

%
\subsubsection{Do I need to define tadpoles?}
If tadpoles are allowed by symmetry we encourage the user to define them --
even if they are absent at tree-level.

%
\subsubsection{How do I define Dirac and Majorana masses?}
Fermion masses are defined analogously to scalar masses.
We stress that \dralgo{} only works with Weyl fermions.
For example, assume a theory with
two Weyl fermions $\psi_1$ and $\psi_2$.
The following terms are then possible
\begin{align}
    m_1 \psi_1\psi_1
  + m_2 \psi_2\psi_2
  + \mD\psi_1\psi_2
  + \mathrm{h.c.}
\end{align}
Where the first two terms are Majorana masses, and the third is a Dirac-type mass. 
These masses can be defined in \dralgo{} via
\begin{lstlisting}[language=Mathematica,mathescape=true]
InputInv={{1,1},{True,True}}; (*$\psi_1 \psi_1$*)
MassTerm1=CreateInvariantFermion[Group,RepFermion,InputInv][[1]]//Simplify;
InputInv={{2,2},{True,True}}; (*$\psi_2 \psi_2$*)
MassTerm2=CreateInvariantFermion[Group,RepFermion,InputInv][[1]]//Simplify;
InputInv={{1,2},{True,True}}; (*$\psi_1 \psi_2$*)
MassTerm3=CreateInvariantFermion[Group,RepFermion,InputInv][[1]]//Simplify;
\end{lstlisting}
The mass matrix is given by
\begin{lstlisting}[language=Mathematica,mathescape=true]
FermionMasses=1/2*m1*MassTerm1+1/2*m2*MassTerm2+mD MassTerm3;
$\mu$IJ=GradMassFermion[FermionMasses];
$\mu$IJC=SparseArray[Simplify[Conjugate[$\mu$IJ]//Normal,Assumptions->{m1>0,m2>0,mD>0}]];
\end{lstlisting}
See the {\tt WessZumino.m} notebook for a worked example.

%
\subsubsection{What if a model has $\overline{\Psi}_1 \gamma_5\Psi_2$ terms?}

Since \dralgo{} only uses Weyl fermions,
the inclusion of other fermionic bilinear operators requires some extra work.
One example is the operator 
$\overline{\Psi}_1 \gamma_5\Psi_2$ which
can be expanded in Weyl spinors
\begin{align}
\Psi_1=\begin{pmatrix}
  \psi_1^L\\
  \psi_1^R
\end{pmatrix}, \quad
\Psi_2=\begin{pmatrix}
  \psi_2^L\\
  \psi_2^R
\end{pmatrix}
\;,
\end{align}
such that
\begin{align}
  \overline{\Psi}_1\Psi_2 &=
    \psi_1^R \psi_2^L
  + \psi_1^L \psi_2^R
  \;,\\
  \overline{\Psi}_1\gamma_5 \Psi_2 &=
  - \psi_1^R \psi_2^L
  + \psi_1^L \psi_2^R
  \;.
\end{align}
These terms can then be implemented in \dralgo{} as explained in earlier examples.

%
\subsubsection{My mass matrix is not diagonal.
Is the effective potential still calculable?}\label{sec:NonDiagonal}

Yes,
it is still possible to calculate the effective potential, just not in general.
It is up to the user to diagonalize the mass matrix.
But once done, the user has two choices.
First, diagonalization-matrices can be given to \dralgo{}.
Second, the user can do all the diagonalization themselves and reload the model.
The first option is quick, but becomes protracted if
the diagonalization is complicated or needs to be done perturbatively.

Consider for example the Standard Model.
As before, field-dependent masses can be created via
\begin{lstlisting}[language=Mathematica,mathescape=true]
DefineNewTensorsUS[$\mu$ij,$\lambda$4,$\lambda$3,gvss,gvvv];
$\phi$vev={0,0,0,$\phi$}//SparseArray;
DefineVEVS[$\phi$vev];
PrintTensorsVEV[];
\end{lstlisting}
To diagonalize the vector-bosons, we first need to extract the field-dependent mass-tensor:
\begin{lstlisting}[language=Mathematica,mathescape=true]
MassMatrix=PrintTensorsVEV[];
VectorMass=MassMatrix[[2]]//Normal;
VectorEigenvectors=FullSimplify[
    Transpose[Normalize/@Eigenvectors[VectorMass[[11;;12,11;;12]]]],
    Assumptions->{g1>0,g2>0,$\phi$>0}];
DVRot={{IdentityMatrix[10],0},{0,VectorEigenvectors}}//ArrayFlatten;
DSRot=IdentityMatrix[4];
RotateTensorsUSPostVEV[DSRot,DVRot];
\end{lstlisting}
After this the effective-potential calculation proceeds as before.

%
\subsubsection{Can \dralgo{} handle non-renormalizable operators?}
Currently no.
This features will be implemented in future versions.

%
\subsubsection{How do I define complicated scalar potentials?}

For most cases, scalar quartics can be easily defined using the examples above.
For non-standard representations,
the model-building tools in \dralgo{} might differ by
a basis change from other conventions in the literature.
Let us take an example,
a $\mathrm{SU}(5)$ model with an adjoint scalar.
Commonly this scalar is written as
$\Phi=\Phi^a T^a$ where
$T^a$ are traceless hermitian matrices satisfying
$\tr T^a T^b=\frac{1}{2}\delta^{ab}$.
The most general (quartic) potential is
\begin{align}
  V_1=
      \delta_1 (\tr\Phi^2)^2
    + \delta_2  \tr\Phi^4
    \;.
\end{align}
By default \dralgo{} uses {\tt GroupMath}, which defines its invariants differently to those above.
In fact,
the two scalar quartic operators given in {\tt GroupMath} can be
linear combinations of those above.
Thus,
the result from \dralgo{} would be in the form
\begin{align}
  V_2=
    \lambda_H \text{inv}_1^{ }(\Phi^4)
  + \lambda_S \text{inv}_2^{ }(\Phi^4)
  \;,
\end{align}
where
the invariants
$\text{inv}_{1,2}$ are the output from the {\tt CreateInvariant} command.
Fortunately,
it is quite easy to find the relations between the $V_1$ and the $V_2$ basis.
To do so, first, rewrite everything in tensor form
\begin{align}
V_1=1/4! \lambda_{1,ijkl}^{ }\Phi_i^{ } \Phi_j^{ } \Phi_k^{ } \Phi_l^{ }
  \;,
\end{align}
and likewise for the \dralgo{} output
\begin{align}
V_2=1/4! \lambda_{2,ijkl}^{ }\Phi_i^{ } \Phi_j^{ } \Phi_k^{ } \Phi_l^{ }
  \;.
\end{align}
Since
$\lambda_1$ and
$\lambda_2$ are defined in different bases,
we want to compare invariants.
Here we only need three:
\begin{align}
 \lambda_{ijkl}\delta_{ij} \delta_{kl}
  \;,\quad
 \lambda_{ijkl}\lambda_{ijkl}
  \;,\quad
 \lambda_{ijnm}\lambda_{nmkl}\lambda_{kl ij}
  \;.
\end{align}
Comparing the invariants one finds
\begin{align}
  \lambda_H=\frac{1}{960} \sqrt{\frac{23}{14}} (130 \delta_1+47 \delta_2)
  \;,\quad 
  \lambda_S=-\frac{5}{64} \sqrt{\frac{13}{42}} (2\delta_1-\delta_2)
  \;.
\end{align}
The above procedure works for any representation as long as
the user can write the quartic operator from \dralgo{} in their preferred form.
See {\tt SU5.m} for a concrete example.

%
\subsubsection{Can I use dimensionally reduced theory for calculating nucleation rates?}
Certainly.
The user can use the effective theory and directly find
the bounce in the 3d. To first approximation the nucleation rate is
$e^{-S_\rmi{3d}}$, where
$S_\rmi{3d}$ is the bounce action in the dimensionally reduced theory.
See~\cite{Linde:1981zj} for the original work,
and~\cite{%
  Gould:2021ccf,Ekstedt:2021kyx,Hirvonen:2021zej,Ekstedt:2022tqk}
for recent examples.

%
\subsubsection{Can I study GUT models with \dralgo{}?}
Most GUT models can be readily studied with \dralgo{}.
However, for models with large representations the running-time rises rather rapidly.
The bottleneck is not the complexity of e.g.\ the scalar potential but rather
the size of the representations of particles.
For example, \dralgo{} rapidly handles the general
Pati-Salam,
$\mathrm{SU}(5)$,
3HDM, or
left-right symmetric model --
even with 20--30 free parameters in the potential.
See
{\tt{SU5.m}},
{\tt{3hdm.m}}, and
{\tt{LRSymmetric.m}} 
for explicit implementations.

%
\subsubsection{How much RAM does \dralgo{} require?}

The required RAM is negligible for most models.
However, for models with 50--100 dimensional representations,
the requirements start to rise.
And for said scenario around 1 GB of RAM is required.
If the package drains too much RAM, it is recommended to not include
two-loop contributions.

%
\subsubsection{How should I report possible bugs?}
We are grateful if any bugs are reported.
We kindly ask users to supply both a short description and
a {\tt Mathematica} file describing the error.
These reports and attached files can provided via the Issue Tracker on 
\url{https://github.com/DR-algo/DRalgo}. 

%
\section{Outlook}
\label{sec:outlook}

High-temperature field theory is pestered by large radiative corrections,
which compromises perturbative calculations.
In effect, perturbation theory needs to be reorganized
in terms of thermal resummations.
While at leading order only thermally corrected masses are required,
new contributions to couplings arise at higher orders.
In addition, there are currently several different schemes for incorporating thermal masses --
all depending substantially on the renormalization scale.

As an effective field theory,
dimensional reduction by-passes these issues.
The ultraviolet sector of the theory is controlled by a matching computation and
the infrared sector is resummed to all orders.
Indeed, as discussed in this paper, using a three dimensional EFT unambiguously resums masses and couplings.
Using the framework is simple from a practical standpoint since
three-dimensional theories are super renormalizable.
Concretely,
only mass parameters are RG-scale dependent and
their beta functions are exact
at two-loop level.
This property not only allows for
a straightforward perturbative treatment,
but also provides
an attractive framework to a non-perturbative lattice study of
the 3d EFT
since in the effective theory relations between continuum and lattice parameters are exact
at two-loop level~\cite{Laine:1997dy}.

Hitherto, dimensional reduction has been used sparsely.
With this paper and the associated software,
we aim to change this by automating the EFT construction.
In summary, \dralgo{} calculates
all effective couplings and masses in the effective theory,
the leading-order beta functions both in the 4d parent and 3d effective theory,
as well as
the effective potential within the EFT.
This facilitates studying models with dimensional reduction and
requires merely 
three-dimensional calculations that are analogous to zero-temperature computations.
Since the effective theory is fully bosonic,
the perturbative calculation can be compared with lattice simulations.
This has one clear benefit as
a large number of fundamental 4d theories can map into the same effective theory
given the mass hierarchy of the additional scalars.
Not only does
the lattice provide an invaluable cross-check, but pre-existing simulations can be reused and applied to new BSM theories.

In conclusion,
gravitational waves have opened up a new gateway to the early universe, and particle physics stands at its threshold.
Upcoming experiments are fast approaching both at future
particle colliders and
gravitational wave observatories.
It is therefore important to control theoretical uncertainties
at unprecedented precision.
In this venture,
dimensional reduction is the tool of choice,
and \dralgo{} its harbinger.

%

\section*{Acknowledgements}

We thank
Renato Fonseca and
Johan L{\"o}fgren for illuminating discussions during the inception of this project.
We express our gratitude to 
Lauri Niemi and
Jorinde van de Vis
for testing the {\tt alpha} version of the software and providing invaluable feedback.
We also thank
Oliver Gould,
Joonas Hirvonen,
Johan L{\"o}fgren,
and
Juuso {\"O}sterman
for interesting discussions.
The work of Andreas Ekstedt has been supported by
the Deutsche Forschungsgemeinschaft under the Germany Excellence Strategy -
EXC $2121$ Quantum Universe - $390833306$; and by
the Swedish Research Council, project number VR:$2021$-$00363$.
Philipp Schicho was supported
by the European Research Council, grant no.~725369, and
by the Academy of Finland, grant no.~1322507.
The work of Tuomas V.~I.~Tenkanen has been supported in part
by the National Science Foundation of China grant no.~19Z103010239.

%
\appendix
\renewcommand{\thesection}{\Alph{section}}
\renewcommand{\thesubsection}{\Alph{section}.\arabic{subsection}}
\renewcommand{\theequation}{\Alph{section}.\arabic{equation}}

%
\section{Interfacing \dralgo{} output}
\label{sec:thermo}

In this appendix,
we collect additional material that is not part of the package itself. 

\subsection{From 3d effective theory to thermodynamics}

A possible interface between
the output of \dralgo{}, in the 3-dimensional theory, and
the determination of thermodynamic quantities 
is implemented in {\tt ./examples/ah-thermo.m}.
This implementation comes with a disclaimer:
we present a simplified algorithm suitable for the simple case of
the Abelian-Higgs model (cf.~eq.~\eqref{eq:L:ah}) but this
is not part of the package itself.
We encourage users to implement their own numerical minimisation routines
for thermodynamics optimized for individual models.
The algorithm given here is not gauge invariant and
should be used with discretion. 

The algorithm~\ref{alg:thermo} illustrates our implementation.
\begin{algorithm}
\caption{
\label{alg:thermo}
  The \dralgo{} algorithm output is interfaced in
  {\tt ./examples/ah-thermo.m} with functions
  {\tt solveBetas[]},
  {\tt DRstep[]} and
  {\tt Veff3d[]}, which are called by
  {\tt findThermo[]}.
}
\begin{algorithmic}
\State {\bf Input:}
  Four-dimensional theory parameters $c_{i} \in \{c_{1},\dots,c_{n}\}$ and
  scale factor $X$
\For{{\bf all}  $c_{i} \in \{c_{1},\dots,c_{n}\}$}
  \Call{{\tt findThermo[$X$,$c_i$]}}
    \State {\bf Call}
        $c_i(\bar{\mu})$ = {\tt solveBetas[$c_i$]}
    \For{{\bf all}  $T_{i} \in \{T_{\rmi{min}},\dots,T_{\rmi{max}} \}$} 
      \Comment{e.g.\ binary search}
      \State Fix $T$-dependent RG-scale: $\bar{\mu} = X\pi T$ 
      \State {\bf Call}
        $c_{i,\rmi{3d}}$ = {\tt DRstep[$T$,$\bar{\mu}$,$c_i$]}
      \State
        Minimize the effective potential by
        {\tt NMinimize[Re[Veff3d[$\phi$,$c_{i,\rmi{3d}}$]]]}
    \EndFor
\State
  Return \{$\Tc^{ }$, $\phi_{\rmi{c}}^{ }/\Tc^{ }$, $L/\Tc^4$\},
  based on degenerate minima.
\EndCall
\EndFor
\State {\bf Output:}
Thermodynamics as function of $\{c_{1},\dots,c_{n}\}$
\State Export the data
\State Plot by Python: {\tt .examples/ah-thermo-python-plots/plot.py}
\end{algorithmic}
\end{algorithm}
In the simplified algorithm {\tt ah-thermo.m}
the loop over 4-dimensional variables $c_i$ contains only a single $\lambda$ and
for simplicity and faster running time
$M=100$ and
$g^2=0.42$ are fixed.
Here,
$M$ is the Higgs mass in arbitrary units related to
the \MSbar{} mass parameter by
the tree-level relation $\mu^2 = -M^2/2$. 
The critical temperature $\Tc$ is defined from the condition that
the effective potential at the broken minimum vanishes since
the potential at the symmetric minimum at the origin is normalized to zero. 
To find this condition, we use an elementary binary search.

As an indicator of the transition strength, we also output
the value of the background field at the critical temperature,
$\phi_\rmi{c}/\Tc$, that describes the discontinuity of the minima at $\Tc$.
Since $\phi_\rmi{c}/\Tc$ is gauge-dependent, it should
not be given physical meaning but 
used as a rough indicator that correlates positively with
the phase transition strength.
This strength can be defined in terms of
the released latent heat at the critical temperature $\Tc$ 
\begin{equation}
L = T \Delta p' =
  T^{ }\Delta \frac{{\rm d}V^{\rmi{4d}}_{\rmi{eff}}}{{\rm d}T} =
  T^{2}\Delta \frac{{\rm d}V^{\rmi{3d}}_{\rmi{eff}}}{{\rm d}T}
  \;,
\end{equation}
where
$p'$ is the derivative of the pressure with respect to temperature and
$\Delta \equiv
(\dots)_{\rmi{low-}T} -
(\dots)_{\rmi{high-}T}$ denotes the difference between the broken and symmetric phase.
The $T$-derivative is approximated as a finite difference ${\rm d}T = 0.1$.
The symmetric-phase part does not contribute here since we normalized
the effective potential to zero at all $T$ at the origin.
For a commented documentation of the technicalities of this implementation
see directly the source {\tt ah-thermo.m}.       

The described perturbative determination is done in Landau gauge
with $\xi = 0$ in eq.~\eqref{eq:prop:gauge} and
the results for $\Tc$ and $L/\Tc^4$ are gauge-dependent.
Despite this simple user-friendly example,
a recipe for a more sophisticated, gauge-invariant determination can be found
in e.g.~\cite{Schicho:2022wty} (cf.\ refs.\ therein).
The final output data for thermodynamics is\\
stored in
{\tt ./examples/ah-thermo-python-plots/*.dat},\\ 
plotted in {\tt ./examples/ah-thermo-python-plots/plot.py}, and
shown in fig.~\ref{fig:ah-thermo}.
\begin{figure}[t]
\centering
\includegraphics[width=0.32\textwidth]{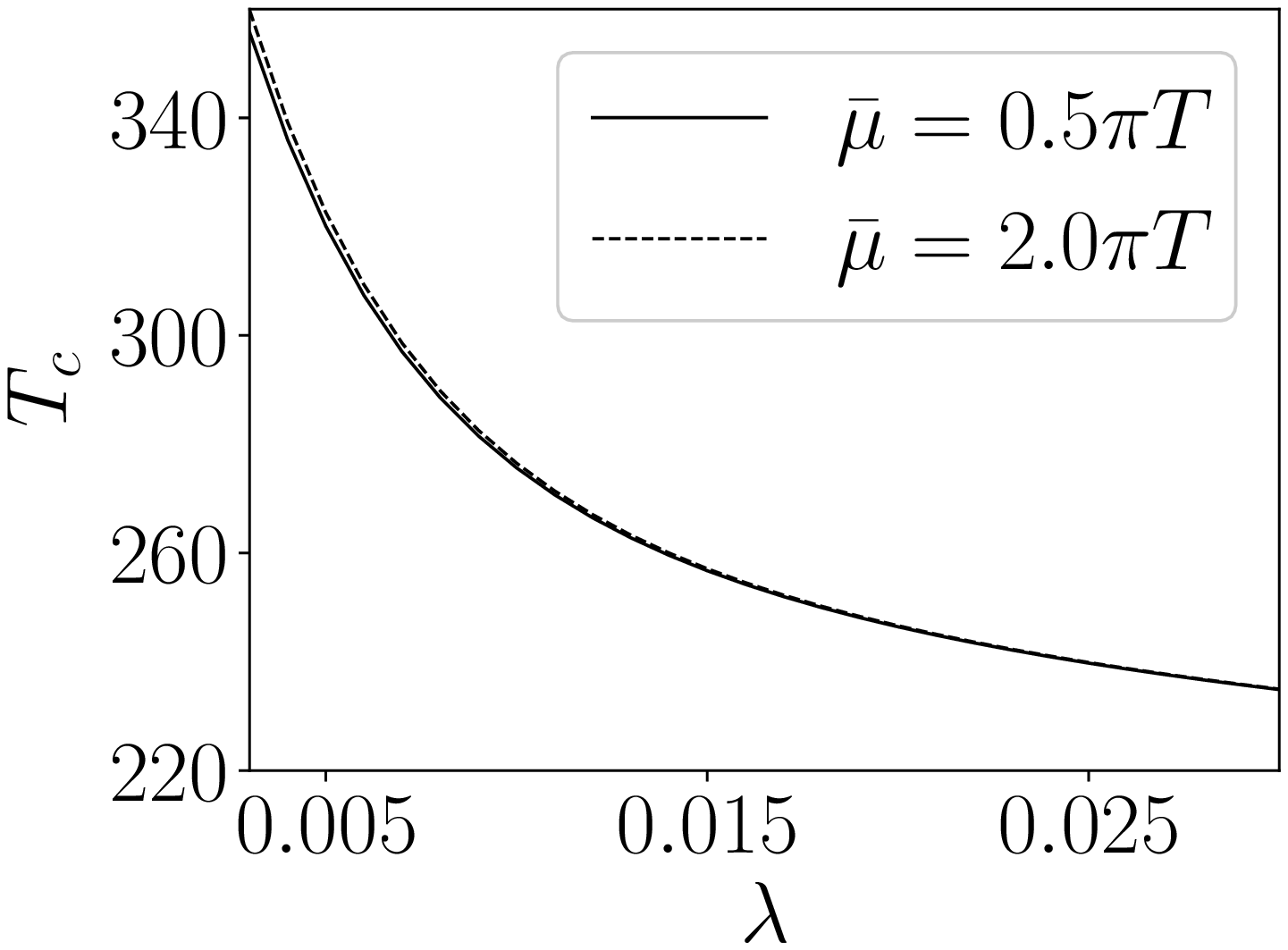} 
\includegraphics[width=0.32\textwidth]{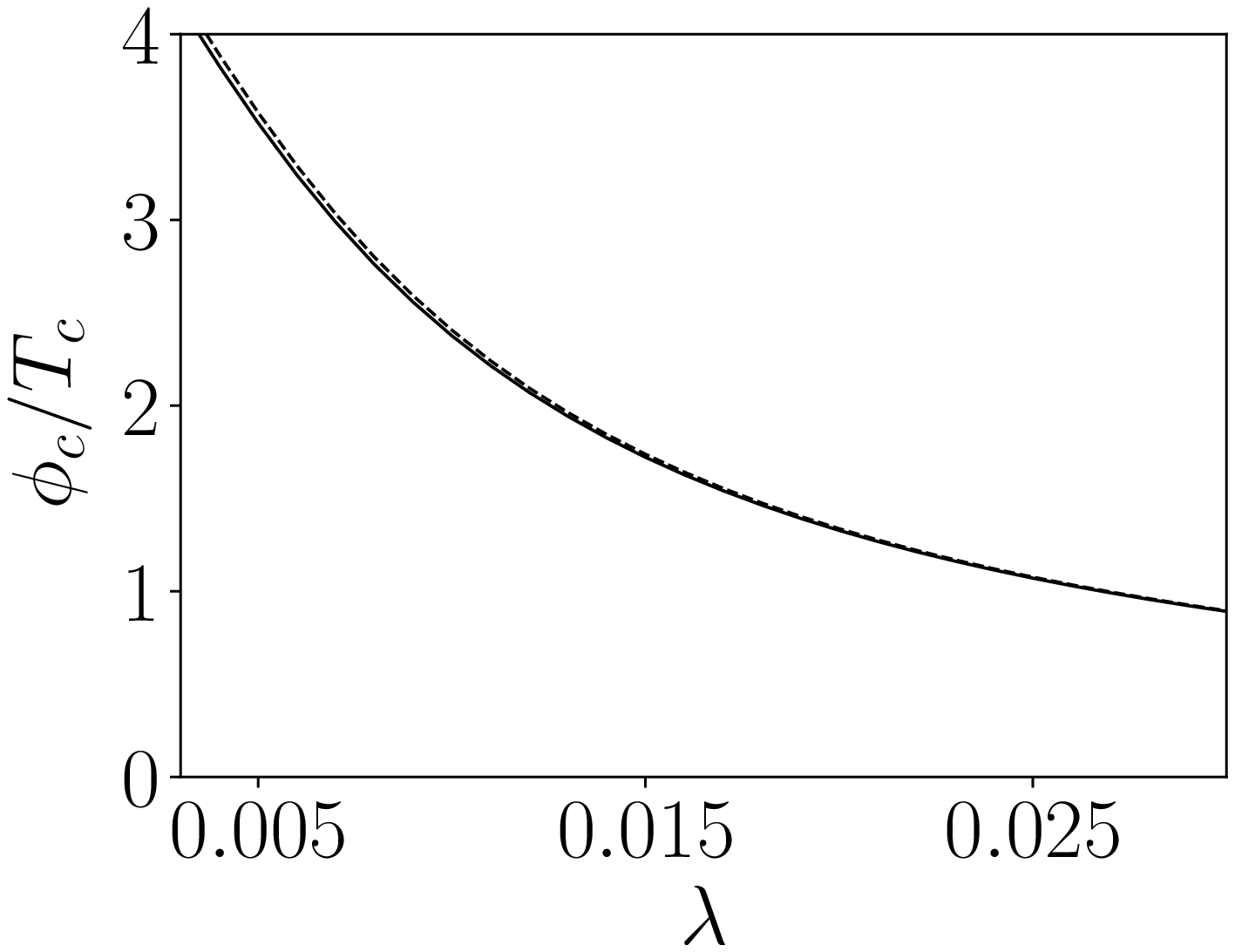} 
\includegraphics[width=0.32\textwidth]{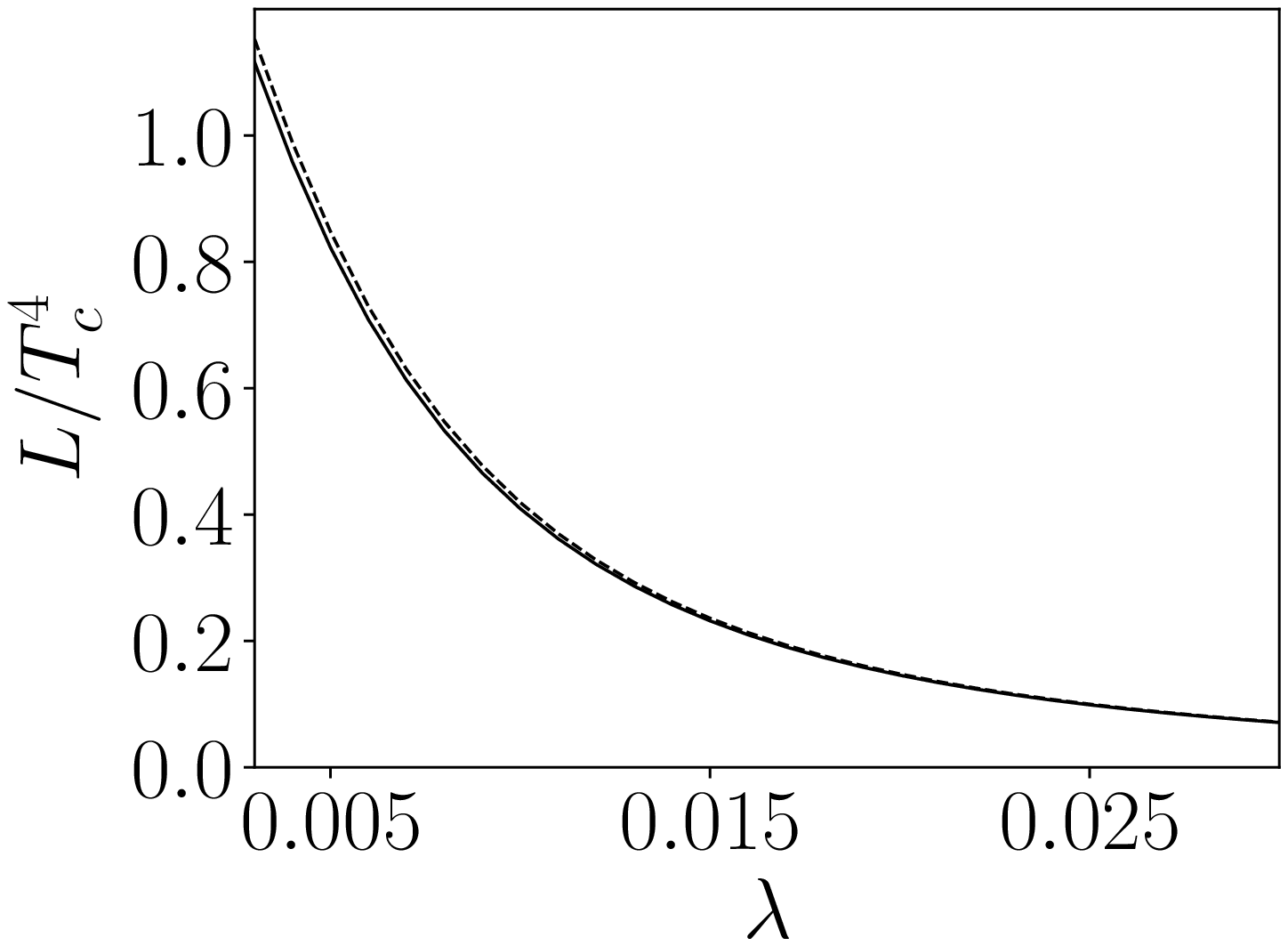}  
\caption{%
  Thermodynamics of
  the Abelian-Higgs model as function of $\lambda$ (cf.~eq.~\eqref{eq:L:ah})
  at fixed
  $g^2=0.42$ and
  $M=100$
  as obtained in
  {\tt ./examples/ah-thermo.m}.
  Both $M$ and $T$ are in arbitrary units of mass.
  The RG-scale is varied at the edges of the interval
  $\bmu = (0.5\dots2.0)\pi T$.
  While the curves are barely discernible,
  this showcases that the scale-dependence is indeed a higher-order effect
  once at full NLO dimensional reduction.
  The additional RG-scale of the 3d EFT is not varied in the plots.  
}
\label{fig:ah-thermo}
\end{figure}

We encourage \dralgo{} users to develop and optimize their
individual implementations for algorithms to determine thermodynamic properties.
While
including efficient algorithms for determining thermodynamics is conceivable
for future versions of \dralgo{},
in the current version \DRalgoVersion{}
these features are not implemented in the package itself.

{\small
%

}
\end{document}